\documentclass[journal]{IEEEtran}
\usepackage{xcolor}
\usepackage{amssymb, amsmath, bbm, graphicx}
\usepackage[inline]{enumitem}
\usepackage[linesnumbered,ruled,vlined]{algorithm2e}
\usepackage{amsthm}
\usepackage{multirow}
\usepackage[capitalise]{cleveref}
\usepackage[subrefformat=parens,labelformat=parens]{subfig}
\usepackage{array}
\usepackage[T1]{fontenc}   % This is used to encode some Chinese characters, especially for Chinese author names

\theoremstyle{remark}
\newtheorem{proposition}{Proposition}

\theoremstyle{remark}

\theoremstyle{remark}

\newcommand{\cmnt}[1]{\ignorespacesafterend}
\newcommand{\Power}{30}
\newcommand{\Du}{250}
\newcommand{\figW}{0.50}

\newcommand{\cdim}[2]{\mathbb{C}^{#1 \times #2}}

\newcommand{\setR}[1]{\mathcal{R}_{#1}}
\newcommand{\setK}[1]{\mathcal{K}_{#1}}
\newcommand{\myFrac}[2]{\genfrac{}{}{0pt}{}{#1}{#2}}
\newcommand{\rate}[2]{\mathcal{R}_{#1, #2}}
\newcommand{\diag}[1]{\text{diag}\left(#1\right)}

\usepackage[
backend=biber,
style=ieee,
sorting=none
]{biblatex}
     
\let\oldnl\nl% Store \nl in \oldnl
\newcommand{\nonl}{\renewcommand{\nl}{\let\nl\oldnl}}% Remove line number for one line

\graphicspath{ {./figures/} }
\DeclareGraphicsExtensions{.eps,.pdf,.png}
\DeclareMathOperator{\Tr}{Tr}
\DeclareMathOperator{\evd}{evd}
\DeclareMathOperator{\vectrz}{vec}

\begin{document}

\title{Hybrid Precoding and Combining for O-RAN Massive MIMO Systems}

%authors
\author{Seraphin~F.~Kimaryo,~\IEEEmembership{Member,~IEEE}, Jung-Mo~Moon, and Kyungchun~Lee,~\IEEEmembership{Senior Member,~IEEE}% <-this % stops a space
    \thanks{This work was supported by Institute of Information \& communications Technology Planning \& Evaluation (IITP) grant funded by the Korea government (MSIT) (RS-2024-00396828, Development of AI based low power 5G-A O-DU/O-CU). (Corresponding author: Kyungchun Lee.)}% <-this % stops a space
    \thanks{S. F. Kimaryo and K. Lee are with the Department of Electrical and Information Engineering and the Research Center for Electrical and Information Technology, Seoul National University of Science and Technology, Seoul 01811, Republic of Korea (e-mails: (seraphin, kclee)@seoultech.ac.kr).}
    \thanks{J.-M. Moon is with the Intelligent RAN SW Research Section, Electronics and Telecommunications Research Institute (ETRI), 218 Gajeong-ro, Daejeon, 34129, Republic of Korea (e-mail: jmmoon@etri.re.kr).}
}

% Make the title area
\maketitle

% Abstract
\begin{abstract} Recently, the radio access network (RAN) has been reimagined through open RAN (O-RAN) architecture to overcome the limitations of the traditional monolithic, black-box design philosophy and to leverage emerging technologies such as cloud computing, machine learning, and artificial intelligence. Although O-RAN disaggregates the base station (BS) into independent components to enable modular deployment, this paradigm introduces new challenges in signal processing and beamforming. In particular, beamforming in O-RAN must account for fronthaul link (FHL) capacity constraints, coordinated operation of multiple radio units (possibly from different vendors), and resource management, including transmit power allocation.
To address these challenges, this work proposes novel beamforming techniques for a downlink multiuser hybrid massive multiple-input multiple-output system, in which hybrid architectures are deployed at both the users and at each BS radio unit. We formulate a sum-rate maximization problem incorporating transmit-power, unit-modulus, and FHL capacity constraints. The proposed algorithm integrates gradient descent and the Lagrange multiplier method to design analog beamformers and digital precoders, respectively, whereas digital combiners are designed to render the received streams parallel to each other. By avoiding computationally intensive operations such as eigenvalue and singular value decompositions, the proposed approach, as evidenced by simulation results, achieves superior performance with moderate computational complexity compared to existing benchmark schemes.
\end{abstract}

\IEEEpeerreviewmaketitle

\section{Introduction}
% The very first letter is a 2-line initial drop letter followed
\IEEEPARstart{T}{he} traditional radio access network (RAN) has long functioned as a black box to all stakeholders except equipment vendors. This lack of transparency is widely regarded as a major bottleneck to advanced network optimization, limiting achievable performance gains and cost reductions. Moreover, the closed nature of traditional RANs enables vendor lock-in, whereby network operators are compelled to procure complete RAN solutions from a single vendor, thereby precluding interoperability among solutions from different providers.

Open RAN (O-RAN) represents a framework for transforming conventional RANs into open, intelligent, virtualized, and fully interoperable systems. This is achieved by standardizing and opening RAN interfaces to enable multi-vendor interoperability, integrating cloud-computing resources, and incorporating RAN Intelligent Controllers (RICs) \cite{10024837}. As such, O-RAN introduces a new paradigm for rearchitecting the RAN and provides a native platform for integrating recent advances in machine learning (ML) and artificial intelligence (AI) into RAN operations.

The O-RAN architecture is the outcome of the O-RAN alliance initiative, a global consortium of mobile network operators, vendors, research organizations, and academic institutions. The alliance seeks to reshape the RAN ecosystem toward more intelligent, open, virtualized, and fully interoperable mobile networks.
As noted earlier, the core philosophy of O-RAN is to enable an open and interoperable RAN, allowing hardware and software solutions from multiple vendors to operate seamlessly together. To this end, O-RAN is designed around the following key principles:
\subsubsection{Virtualization}
Under this principle, the software is deliberately separated from hardware, whereby both components are developed in accordance with standardized specifications to ensure interoperability among solutions from different vendors. This process, commonly referred to as virtualization and primarily addressed by O-RAN working group 6, includes the deployment of cloud networks, which are collections of computing resources located at one or multiple sites, to support the execution of RAN-related software functions. Consequently, multiple nodes and/or tenants can share the same computing infrastructure, thereby increasing deployment flexibility and reducing operational expenditures.

\subsubsection{Disaggregation}
To further enhance flexibility, base station (BS) hardware is decomposed into three functional entities: the open central unit (O-CU), the open distributed unit (O-DU), and the open radio unit (O-RU) \cite{10024837}. The O-RU, which operates as a leaf node, typically comprises the antenna system and performs low-layer signal processing functions, including precoding, beamforming, digital-to-analog and analog-to-digital conversion, and fast Fourier transforms. The O-DU primarily handles medium access control and physical layer processing functions, such as modulation and scrambling, whereas the O-CU is responsible for higher-layer functionalities, including radio resource control, quality of service management, and security-related operations.

\subsubsection{Introduction of RAN intelligent controllers (RICs)}
To enable intelligent RAN operation and facilitate the integration of emerging ML and AI techniques, the O-RAN architecture introduces RAN intelligent controllers (RICs). These logical, programmable components perform control and optimization functions, such as traffic-aware network optimization, scheduling, handover management, and load balancing. Depending on the latency requirements of a given function, it is assigned to either a non-real-time (non-RT) RIC, which supports control loops with timescales exceeding 1 s, or a near-real-time (near-RT) RIC, which supports control loops operating on timescales from 10 ms to 1 s.
Accordingly, non-RT RICs are typically deployed within the service management and orchestration layer to support high-level network management tasks, including policy control and AI/ML model management. In contrast, near-RT RICs are deployed closer to the RAN edge to support latency-sensitive functions, such as mobility management and load balancing.

\subsubsection{Opening and standardization of interfaces}
Following the disaggregation of hardware and the introduction of logical RAN nodes, standardized interfaces are required to enable efficient information exchange and control signaling among these components. This brings us to the interfacing part. To this end, the O-RAN architecture defines, opens, and standardizes a set of interfaces to ensure seamless interoperability among multi-vendor equipment. A non-exhaustive list of such interfaces includes the A1 interface, which connects the non-RT RIC and near-RT RIC; the E2 interface, which links the near-RT RIC to RAN edge nodes, such as the O-CU and O-DU; the F1 interface between the O-CU and O-DU; and the fronthaul link (FHL), which connects the O-RU and O-DU \cite[Fig. 1]{11165760}.

Collectively, these design principles enhance RAN configurability and flexibility, enable the integration of ML- and AI-driven optimization, embrace a white-box design philosophy, mitigate vendor lock-in, lower entry barriers for emerging vendors offering specialized solutions, and foster competition, which is a key driver of innovation and performance improvement.

The aforementioned advantages have attracted significant interest not only from industry \cite{TELUSPartnersSamsung, samsungTELUS} but also from academia and research institutions. Recently, \cite{11124199} surveyed the network-slicing framework in O-RAN, discussing deployment scenarios, representative use cases, and interactions among different slices and other components of the cellular network, including the core and transport networks. Notably, \cite{10024837} provides a comprehensive overview of the O-RAN architecture, interfaces, security considerations, and open challenges, whereas \cite{10439167} focuses on the intelligent aspects of O-RAN by examining the adoption of AI and ML models for near-RT and non-real-time non-RT RIC operations.

Owing to the openness of O-RAN interfaces, security threats have emerged as a critical concern. To address these challenges, \cite{10606000} analyzed the resilience of several O-RAN interfaces under denial-of-service attacks, evaluated the resulting performance degradation, and documented the associated risks, vulnerabilities, and recommended mitigation strategies. Similarly, the authors of \cite{10433004} identified significant access-control vulnerabilities in O-RAN xApps through systematic security testing, demonstrating their potential to disrupt RAN operations. These vulnerabilities, along with recommended security enhancements, were subsequently reported to the O-RAN Alliance.

By leveraging pedestrian flow analytics and the capacity of non-terrestrial networks, \cite{10521586} proposes an energy-efficient power management architecture for O-RAN-based BSs. In a related effort, \cite{10380322} adopts the O-RAN architecture to analyze a cell-free (CF) multiple-input multiple-output (MIMO) system, investigating the performance of two functional split options and various resource orchestration strategies. Given its central role in O-RAN architecture, orchestration was further examined in \cite{10835138}. Moreover, \cite{10330597} investigated a decentralized pilot-assignment strategy for O-RAN-based CF massive MIMO systems by training multiple deep reinforcement learning agents to mitigate pilot contamination. User association with O-RUs, as well as O-CU and O-DU placement strategies in O-RAN deployments, are studied in \cite{9946423}, where deep reinforcement learning is employed to minimize deployment cost and end-to-end average user delay.

Although the aforementioned works addressed several critical aspects of O-RAN architecture, only a limited number of studies investigated beamforming, particularly under interface capacity constraints and multi–O-RU coordination. Beamforming is a fundamental signal processing technique that enables MIMO systems to efficiently exploit multiple antennas for both transmission and reception. In this context, this work focuses on beamforming for hybrid MIMO systems (HMSs) \cite{8371237}. HMSs are well known for mitigating the rapid growth in the number of radio-frequency (RF) chains as antenna counts increase, by employing low-power phase shifters to map signals between a limited number of RF chains and a large antenna array. This architecture significantly reduces transceiver power consumption. Depending on the interconnection structure of the phase shifters, HMSs can be classified as fully connected or partially connected \cite{8371237}, which gives rise to multiple design trade-offs and application scenarios.

Early efforts in O-RAN beamforming include \cite{electronics10172162} and \cite{en15072429}, which investigated zero-forcing (ZF) and minimum mean-square error (MMSE) precoding for downlink multiple-input single-output (MISO) systems, respectively. While \cite{electronics10172162} proposes ZF precoding for general-purpose fifth-generation (5G) O-RAN deployments, \cite{en15072429} tailors ZF and MMSE techniques to Internet-of-Energy applications. However, these studies primarily focused on fully digital architectures and do not consider HMSs.

Addressing this gap, recent works \cite{11165760,11028877} investigated hybrid beamforming for O-RAN-based millimeter-wave (mmWave) massive MIMO systems. Specifically, \cite{11165760} proposed a non–grid-of-beams (non-GoB) approach for analog precoder design and employed successive convex approximation and CVX to optimize digital precoders. To alleviate the resulting computational burden, a deep learning (DL)-based solution was introduced. The framework in \cite{11165760} was subsequently extended in \cite{11028877} by exploiting statistical channel state information to reduce information exchange over the FHL.

Despite these advances, there remains considerable room for improving both system performance and computational efficiency. Towards this objective, this work introduces algorithmic designs that simultaneously enhance performance and reduce computational complexity while satisfying all hardware and resource constraints. In contrast to \cite{11165760}, which optimizes analog precoders and combiners only once to reduce complexity, this work prioritizes performance by iteratively updating both precoders and combiners using low-complexity procedures. Moreover, whereas prior approaches rely heavily on computationally intensive operations such as singular value decomposition (SVD), the proposed methods in this work avoid such operations without compromising performance, thereby achieving further complexity reduction.

Furthermore, in the design of analog beamformers, the non-convex unit-modulus constraint imposed by phase shifters presents a significant challenge, not only in achieving near-optimal performance but also in obtaining a tractable solution. Consequently, most existing studies adopt codebook-based techniques, such as orthogonal matching pursuit (OMP) \cite{6717211, 8048606, 7878572}, approximation-based approaches such as non-GoB methods \cite{11165760, 11028877}, or highly relaxed formulations in which analog beamformers are designed using local channel state information \cite{9747002, 7160780}. Although these approaches are computationally tractable, they typically yield suboptimal performance.
Conversely, this work designs the analog beamformers using a simple gradient-descent (GD)-based method applied to a reformulated, nonrelaxed problem, whose objective function and its gradient can be computed with low complexity.

This is achieved through a two-step process. First, we adopt the MMSE approach \cite{11237065,5756489}, which introduces auxiliary variables to reformulate the objective function into a highly tractable form that exhibits convexity with respect to certain optimization variables. Next, we exploit the mixed-product property of the Kronecker product to eliminate all EVDs, which are initially used to decompose the objective function for analog precoder/combiner optimization. Since EVDs are computationally expensive, their elimination results in a low-complexity solution. Note that, because this work and prior works \cite{11165760,11028877} both investigate hybrid beamforming in O-RAN architectures, their system models and some constraints have similar forms. 
The primary differences, however, lie in the proposed solution approaches. As discussed above, this work employs GD for analog precoder and combiner design, the Lagrange multiplier method for digital precoder design, and a digital combiner design that promotes parallelism among the received streams. Collectively, these approaches enable the proposed algorithm to achieve a favorable performance-complexity tradeoff relative to the considered benchmark schemes.
In contrast, \cite{11165760} and \cite{11028877} employ CVX and SVD for digital precoder and combiner design, respectively, while their analog precoders and combiners are designed using non-GoB and Riemannian manifold optimization techniques, respectively. Thus, the main contributions of this work are summarized as follows:
\begin{itemize}
    \item This work proposes robust beamforming techniques for fully connected HMSs within the O-RAN architecture. Unlike existing approaches, the proposed methods jointly prioritize sum-rate maximization and computational complexity reduction while satisfying all practical hardware and resource constraints.
    \item In the quest to achieve sum-rate maximization, we observe that many existing techniques adopt performance-degrading strategies in the design of analog beamformers, such as problem relaxation, simplistic approximations, and codebook-constrained solutions. Avoiding these strategies is essential to unlock additional performance gains. Motivated by its simplicity and robustness, this work employs GD to design the analog beamformers at each O-RU and user. Accordingly, we develop novel formulations that not only avoid the aforementioned performance-degrading procedures but also render the resulting subproblems highly tractable to GD-based optimization.
    Furthermore, although subtle, we avoid the common relaxation of excluding the analog precoder from the transmit power constraint during digital precoder design. Collectively, these design choices lead to a substantial improvement in the achievable system sum rate.
    \item In addition, most existing schemes, such as \cite{11165760, 11028877, 8048606, 9747002}, rely heavily on computationally intensive operations, including SVD and eigenvalue decomposition (EVD). In contrast, this work proposes an innovative approach that leverages the mixed-product property of the Kronecker product to circumvent these operations, thereby significantly reducing computational complexity. Quantitative evaluations demonstrate that eliminating such operations reduces the computational complexity of the proposed algorithm by more than 99\%.
    \item Extensive numerical results are presented to compare the proposed framework with state-of-the-art benchmark schemes. The results demonstrate that the proposed approach achieves superior performance with moderate computational complexity. Specifically, against the best-performing benchmark scheme, the proposed algorithm maintains a performance gain of at least 50 bps/Hz across the tested configurations, increasing to over 134 bps/Hz when the number of transmit antennas is large.
\end{itemize}

\textit{Paper organization}: System model and problem formulation are presented in \cref{Sect:sys_model}. The proposed solution to the formulated problem is presented in \cref{sect:analytical_soln}. \cref{Sect:Simulations} presents numerical results, whereas \cref{Sect:Concl} concludes the paper.

\textit{Notations}: $\cdim{m}{n}$ denotes an $m \times n$ set of complex numbers, whereas $\Re\left( \cdot \right)$ denotes the real part of the complex-valued parameter. The superscripts $*$, $H$, and $T$ denote the complex conjugate, Hermitian, and transpose, respectively. Vectors and matrices are denoted by bold-face lower-case and upper-case letters, respectively, whereas scalars are denoted by italic lower-case letters. $\mathbb{E}$ denotes the statistical expectation operator. The inverse, trace, Frobenius norm, EVD, and vectorization of a matrix $\boldsymbol{A}$ are denoted by $\boldsymbol{A}^{-1}$, $\Tr\left(\boldsymbol{A}\right)$, $\| \boldsymbol{A} \|_{F}$, $\evd\left( \boldsymbol{A} \right)$, and $ \vectrz\left( \boldsymbol{A} \right)$, respectively. Similarly, the operator $\left|\cdot\right|$ denotes a matrix determinant, absolute value, or cardinality if its argument is a matrix, scalar, or a set, respectively. 
$\left[ \boldsymbol{A} \right]_{m,n}$ denotes the element at the $m$-th row and $n$-th column of matrix $\boldsymbol{A}$. The Euclidean norm and the $n$-th element of a vector $\boldsymbol{a}$ are denoted by $\| \boldsymbol{a} \|$ and $\left[ \boldsymbol{a} \right]_{n}$, respectively. $\boldsymbol{I}$ denotes an identity matrix of compatible dimensions, whereas $\diag{x_{1}, \cdots, x_{n}}$ denotes a diagonal matrix whose diagonal elements are $x_{1}, \cdots, x_{n}$.

\section{System Model and Problem Formulation} \label{Sect:sys_model}
\subsection{Parameters definition and achievable rate formulation}
In this work, we analyze the spectral efficiency (SE) performance of a single-cell downlink multiuser (MU) hybrid MIMO system. Specifically, we consider an O-RAN deployment in which a BS, disaggregated into a single O-DU and $R$ O-RUs, serves $K$ users randomly located within a given coverage area. Following BS disaggregation, each O-RU $r$ is connected to the O-DU via an FHL with a maximum capacity/throughput of $FL_{r}$ [bps]. Moreover, since O-RUs in O-RAN networks may be either co-located or geographically distributed, let $D_{r,u}$ denote the distance between O-RU $r$ and user $u$. Each O-RU is equipped with a hybrid transceiver architecture comprising $M_{t}$ RF chains, fully connected to $N_{t} > M_{t}$ antennas via $N_{t}M_{t}$ phase shifters. This configuration is replicated across users, where the numbers of RF chains, antennas, and phase shifters at each user are denoted as $M_{r}$, $N_{r}$, and $N_{r}M_{r}$, respectively, where $N_{r} > M_{r}$. To serve user $u$, O-RU $r$ uses a digital and analog transmit precoders $ \boldsymbol{F}_{BB,r,u} \in \mathbb{C}^{M_{t} \times \Pi_{r,u}}$ and $\boldsymbol{F}_{RF,r} \in \mathbb{C}^{N_{t} \times M_{t}}$, respectively, to precode a unit-power data symbol $\boldsymbol{d}_{r,u} \in \mathbb{C}^{\Pi_{r,u} \times 1}$, where $\Pi_{r,u}$ denotes the number of data streams allocated to user $u$ by O-RU $r$. For algorithm stability and system performance reliability, these streams are constrained such that $\sum_{r \in \setR{u}} \Pi_{r,u} \leq M_{r}, \forall u$.
The transmitted signal of O-RU $r$ is obtained by aggregating the precoded symbols of all served users as $ \boldsymbol{F}_{RF,r} \sum_{u \in \mathcal{K}_{r}} \boldsymbol{F}_{BB,r,u} \boldsymbol{d}_{r,u}$, and is transmitted over the channels $ \boldsymbol{H}_{r,u} \in \cdim{N_{r}}{N_{t}}, u \in \mathcal{K}_{r}$ subject to a maximum transmit power $P_{r}$. Here, $\mathcal{K}_{j}$ denotes the set of users served by O-RU $j$ and $\boldsymbol{H}_{j,i}$ represents the channel between O-RU $j$ and user $i$. In the absence of noise, the signal received at the antenna array of user $u$ is given by
\begin{equation}
    \boldsymbol{y}_{u} =  \sum_{r =1}^{R} \boldsymbol{H}_{r,u} \boldsymbol{F}_{RF,r} \sum_{i \in \mathcal{K}_{r}} \boldsymbol{F}_{BB,r,i} \boldsymbol{d}_{r,i}.
\end{equation}
At the receiver, user $u$ first applies an analog combiner  $\boldsymbol{W}_{RF,u} \in \cdim{M_{r}}{N_{r}}$ to map the received signals from the antenna elements to the RF chains. This is followed by a set of digital combiners $\boldsymbol{W}_{BB,u,r} \in \cdim{\Pi_{r,u}}{M_{r}}, \forall r \in \setR{u}$, to recover the data streams transmitted from its serving O-RUs, where $\setR{u}$ denotes the set of O-RUs serving user $u$. The resulting signal processing operation can be expressed as
\begin{equation}
    \hat{\boldsymbol{d}}_{u} = \boldsymbol{W}_{BB,u} \boldsymbol{W}_{RF,u} \left( \boldsymbol{y}_{u} +  \boldsymbol{z}_{u} \right),
\end{equation}
where $\boldsymbol{W}_{BB,u} = \left[ \boldsymbol{W}_{BB,u,1}^{T},\cdots,\boldsymbol{W}_{BB,u,\left|\setR{u}\right|}^{T} \right]^{T} \in \cdim{\aleph_{u}}{M_{r}}$ and $\aleph_{u} = \sum_{r \in \setR{u}} \Pi_{r,u}$ denotes the total number of data streams intended for user $u$. Furthermore, $\hat{\boldsymbol{d}}_{u}$ denotes the vector of filtered data symbols for user $u$, whereas $\boldsymbol{z}_{u}$ represents user $u$'s additive white Gaussian noise with zero mean and $\sigma_{u}^{2}$ variance. The data symbols $\boldsymbol{d}_{r,u}$ and $\boldsymbol{d}_{j,i}$, as well as the noise $\boldsymbol{z}_{u}$, are assumed to be mutually independent for $r\neq j$ and $u \neq i$.
Consequently, the achievable rate between O-RU $r$ and user $u \in \setK{r}$ is given as \cite{8048606}
\begin{multline} \label{eq:Rate_ru}
    \rate{r}{u} = \log_{2} \Big| \boldsymbol{I} + \boldsymbol{C}_{r,u}^{-1} \boldsymbol{W}_{r,u} \boldsymbol{H}_{r,u} \boldsymbol{F}_{RF,r} \boldsymbol{F}_{BB,r,u} \\
    \times \boldsymbol{F}_{BB,r,u}^{H} \boldsymbol{F}_{RF,r}^{H} \boldsymbol{H}_{r,u}^{H} \boldsymbol{W}_{r,u}^{H} \Big|,
\end{multline}
where $\boldsymbol{C}_{r,u} = \sigma_{u}^{2} \boldsymbol{W}_{r,u} \boldsymbol{W}_{r,u}^{H} + \boldsymbol{W}_{r,u} \sum_{l \neq r}^{R} \boldsymbol{H}_{l,u} \boldsymbol{F}_{RF,l} \sum_{i \in \setK{l}} \boldsymbol{F}_{BB,l,i} \boldsymbol{F}_{BB,l,i}^{H}  \boldsymbol{F}_{RF,l}^{H} \boldsymbol{H}_{l,u}^{H} \boldsymbol{W}_{r,u}^{H} \\+ \boldsymbol{W}_{r,u} \boldsymbol{H}_{r,u} \boldsymbol{F}_{RF,r} \sum_{ \myFrac{j \in \setK{r}}{j \neq u} } \boldsymbol{F}_{BB,r,j} \boldsymbol{F}_{BB,r,j}^{H} \boldsymbol{F}_{RF,r}^{H} \boldsymbol{H}_{r,u}^{H} \boldsymbol{W}_{r,u}^{H}$ and $\boldsymbol{W}_{r,u} = \boldsymbol{W}_{BB,u,r} \boldsymbol{W}_{RF,u}$.

\subsection{FHL and its effect}
Notice that the SE in (\ref{eq:Rate_ru}) is expressed solely in terms of the air-interface channels and the design parameters (i.e., precoders and combiners). However, in distributed MIMO systems such as O-RAN, the capacity of the FHL plays a crucial role in determining whether the efficiency in (\ref{eq:Rate_ru}) can be fully realized. This is because the data bits to be transmitted by each O-RU and/or the designed parameters must be delivered from the O-DU to the O-RUs through the FHL. Consequently, the FHL can become a bottleneck if it does not provide sufficient capacity.

This paper adopts category A\footnote{Because precoding is performed at the O-DU in category A, the O-RU requires less complex hardware, consumes less power, and is consequently lightweight and portable. In contrast, the category-A 7-2x functional split requires a high-throughput FHL to accommodate the large volume of information exchanged between the O-DU and O-RUs.} of the O-RAN 7-2x functional split \cite{10024837}, in which the O-DU designs all parameters (i.e., precoders and combiners) and shares them, together with the data bits, with each O-RU. Among the several strategies for transmitting this information over the FHL, this work adopts the compress-and-precode strategy \cite[(11)]{10937259}. Specifically, the O-DU quantizes each element of the designed parameters, represents each quantization level using a codeword with a given bit width \cite{10478417}, and then transfers the resulting bits, together with the data bits, over the FHL. Consequently, the FHL capacity of O-RU $r$ must satisfy
\begin{equation} \label{eq:FHL_relation}
FL_{r} \geq Q_{r} + O_{r} + \frac{BW \sum_{u \in \setK{r}} \rate{r}{u}}{R_{code}},
\end{equation}
where $O_{r}$ denotes the number of overhead bits\footnote{These bits can be used to define the payload structure and/or indicate the beginning and end of a block of bits representing a single piece of information.}, $BW$ denotes the system bandwidth,
\begin{align}
    & Q_{r} = \nonumber \\
    & \frac{N_{t} M_{t} q_{a} + \sum_{u \in \setK{r}} \big( 2q_{d} M_{t} \Pi_{r,u} + 2q_{d} M_{r} \Pi_{r,u} + N_{r} M_{r} q_{a} \big) }{R_{use} \cdot t_{slot}},
\end{align}
$q_{a}$ denotes the number of bits used to represent the phase of each element of the analog precoder/combiner; $q_{d}$ denotes the number of bits used to represent the magnitude or phase (alternatively, the in-phase or quadrature component) of each element of the digital precoder/combiner; $t_{slot}$ denotes the coherence time during which the channel remains stationary; and $R_{use}$ and $R_{code}$ denote the reuse factor of the designed precoders/combiners and the channel coding rate, respectively. Note that because the elements of the digital precoders and combiners generally have different magnitudes, their in-phase and quadrature components are encoded separately, which accounts for the factor of 2 preceding $q_{d}$.
In $Q_{r}$, it is further assumed that each O-RU serving user $u$ is responsible for feeding back the corresponding digital combiner for that user, whereas the analog combiner can be fed back by any O-RU.

Notice that the last term in (\ref{eq:FHL_relation}) determines the number of data bits representing the actual information for users $u \in \setK{r}$ that are transported over the FHL. Therefore, controlling the SE of the air interface also controls the number of bits transmitted over the FHL. By treating the first two terms as fixed quantities, (\ref{eq:FHL_relation}) can be equivalently expressed as
\begin{equation} \label{eq:FHL_relation2}
\sum_{u \in \setK{r}} \rate{r}{u} \leq \frac{\big(FL_{r} - Q_{r} - O_{r}\big) R_{code}}{BW}.
\end{equation}

\subsection{Problem formulation}
In this work, we aim to maximize the total sum rate of the system by optimizing digital and analog precoders/combiners subject to their respective hardware and resource constraints. This problem is represented as follows:
\begin{subequations}
    \begin{align}
        \mathcal{P}(1): \max _{ \myFrac{\boldsymbol{F}_{BB,r,u}, \boldsymbol{F}_{RF,r},}{\boldsymbol{W}_{BB,u,r}, \boldsymbol{W}_{RF,u}} } \quad & \sum_{r = 1}^{R} \sum_{u \in \setK{r}} \rate{r}{u} \label{eq:P1_obj_fn} \\ 
        \mathrm{s.t.} \quad & \sum_{u \in \setK{r}} \| \boldsymbol{F}_{RF,r} \boldsymbol{F}_{BB,r,u} \|_{F}^{2} \leq P_{r}, \forall r \label{eq:pow_constraint}\\
        & (\ref{eq:FHL_relation2}), \forall r \label{eq:FLr_constraint}\\
        & \left| \left[ \boldsymbol{F}_{RF,r} \right]_{i,j} \right|^2 = 1, \forall_{r,i,j} \label{eq:phase_r_constraint}\\
        & \left| \left[ \boldsymbol{W}_{RF,u} \right]_{i,j} \right|^2 = 1, \forall_{u,i,j}. \label{eq:phase_u_constraint}
    \end{align}
\end{subequations}
In $\mathcal{P}(1)$, the resource constraints in (\ref{eq:pow_constraint}) and (\ref{eq:FLr_constraint}) ensure that the transmit power and FHL capacity of each O-RU do not exceed their respective limits, while the hardware constraints in (\ref{eq:phase_r_constraint}) and (\ref{eq:phase_u_constraint}) guarantee that the phase shifters at the O-RUs and user terminals, respectively, operate within physically feasible ranges.

\section{Proposed Solution to $\mathcal{P}(1)$} \label{sect:analytical_soln}
The joint optimization of all variables in $\mathcal{P}(1)$ is highly challenging due to their strong coupling and inherent non-convexity. To address this difficulty, we show that by optimizing one variable at a time while keeping the remaining variables fixed, $\mathcal{P}(1)$ can be decomposed into a set of convex subproblems. This decomposition enables the application of standard convex optimization techniques for their efficient solution.
To facilitate the transformation into convex subproblems with respect to (w.r.t) each optimization variable, the objective function (\ref{eq:P1_obj_fn}) is reformulated into the following equivalent and more tractable form \cite{11237065, 5756489}:
\begin{align} \label{eq:convenient_obj_fn}
    &\sum_{r = 1}^{R} \sum_{u \in \setK{r}} \rate{r}{u} = \sum_{r = 1}^{R} \sum_{u \in \setK{r}} \log_{2} \left| \boldsymbol{U}_{r,u} \right| + \sum_{r = 1}^{R} \sum_{u \in \setK{r}} \Pi_{r,u} \nonumber \\
    &- \sum_{r = 1}^{R} \sum_{u \in \setK{r}} \Tr\left( \boldsymbol{U}_{r,u} \boldsymbol{V}_{r,u}^{H} \boldsymbol{C}_{r,u} \boldsymbol{V}_{r,u}\right) \nonumber \\
    &- \sum_{r = 1}^{R} \sum_{u \in \setK{r}} \Tr\bigg( \boldsymbol{U}_{r,u} \left( \boldsymbol{I} - \boldsymbol{V}_{r,u}^{H} \boldsymbol{W}_{r,u} \boldsymbol{H}_{r,u} \boldsymbol{F}_{RF,r} \boldsymbol{F}_{BB,r,u} \right) \nonumber \\
    &\times \left( \boldsymbol{I} - \boldsymbol{V}_{r,u}^{H} \boldsymbol{W}_{r,u} \boldsymbol{H}_{r,u} \boldsymbol{F}_{RF,r} \boldsymbol{F}_{BB,r,u} \right)^{H} \bigg),
\end{align}
where
\begin{multline}
   \!\!\!\!\!\!\! \boldsymbol{V}_{r,u} = \Big( \boldsymbol{W}_{r,u} \boldsymbol{H}_{r,u} \boldsymbol{F}_{RF,r} \boldsymbol{F}_{BB,r,u} \boldsymbol{F}_{BB,r,u}^{H} \boldsymbol{F}_{RF,r}^{H} \boldsymbol{H}_{r,u}^{H} \boldsymbol{W}_{r,u}^{H} \\
    + \boldsymbol{C}_{r,u} \Big)^{-1} \boldsymbol{W}_{r,u} \boldsymbol{H}_{r,u} \boldsymbol{F}_{RF,r} \boldsymbol{F}_{BB,r,u}
\end{multline}
and
\begin{equation}
    \boldsymbol{U}_{r,u} = \left( \boldsymbol{I} - \boldsymbol{V}_{r,u}^{H} \boldsymbol{W}_{r,u} \boldsymbol{H}_{r,u} \boldsymbol{F}_{RF,r} \boldsymbol{F}_{BB,r,u} \right)^{-1}
\end{equation}
are the usual MMSE auxiliary variables \cite{5756489}.

\subsection{Optimization of $\boldsymbol{F}_{BB,r,u}$} \label{sect:optimize_Fbb_ru}
Even when all other variables are held constant, the presence and coupling of $\boldsymbol{F}_{BB,r,u}$ in the two resource constraints render its optimization challenging. Nevertheless, the digital precoder offers a high degree of flexibility, as its elements can be designed freely to optimize the objective function and satisfy the problem constraints. Exploiting this property, we temporarily ignore constraint (\ref{eq:FLr_constraint}), i.e., $\sum_{u \in \setK{r}} \rate{r}{u} \leq \infty, \forall r$, which results in the following tractable optimization problem w.r.t $\boldsymbol{F}_{BB,r,u}$:
\begin{subequations}
    \begin{align}
        \mathcal{P}(2): \min _{ \boldsymbol{F}_{BB,r,u} } \quad & -\sum_{r = 1}^{R} \sum_{u \in \setK{r}} \rate{r}{u} \label{eq:P1FBB_obj_fn} \\ 
        \mathrm{s.t.} \quad & \mathrm{(\ref{eq:pow_constraint})}.
    \end{align}
\end{subequations}
Observe that $\mathcal{P}(2)$ can now be solved by the standard Lagrange multiplier method to obtain the following solution:
\begin{equation} \label{eq:FBB_ru_soln}
    \boldsymbol{F}_{BB,r,u} = \left( \mu_{r} \boldsymbol{F}_{RF,r}^{H} \boldsymbol{F}_{RF,r} + \boldsymbol{X}_{r} \right)^{-1} \boldsymbol{A}_{r,u}^{H} \boldsymbol{U}_{r,u},
\end{equation}
 where, $\boldsymbol{A}_{r,i} = \boldsymbol{V}_{r,i}^{H} \boldsymbol{W}_{r,i} \boldsymbol{H}_{r,i} \boldsymbol{F}_{RF,r}$, $\boldsymbol{X}_{r} = \sum_{l = 1}^{R}\sum_{i \in \setK{l}} \boldsymbol{F}_{RF,r}^{H} \boldsymbol{H}_{r,i}^{H} \boldsymbol{W}_{l,i}^{H} \boldsymbol{V}_{l,i} \boldsymbol{U}_{l,i} \boldsymbol{V}_{l,i}^{H} \boldsymbol{W}_{l,i} \boldsymbol{H}_{r,i} \boldsymbol{F}_{RF,r}$, and $\mu_{r}$ is a Lagrange multipler for O-RU $r$, which is chosen by bisection method to satisfy complementary slackness condition $\mu_{r} \left( \sum_{u \in \setK{r}} \| \boldsymbol{F}_{RF,r} \boldsymbol{F}_{BB,r,u} \|_{F}^{2} - P_{r} \right) = 0$.
It is worth mentioning that, after every variable is optimized, $\boldsymbol{F}_{BB,r,u}$ is scaled to meet the original constraint (\ref{eq:FLr_constraint}). To achieve that, we first examine whether the power-compliant precoder, $\boldsymbol{F}_{BB,r,u}$, still violates constraint (\ref{eq:FLr_constraint}), in which case the steps in \cref{prop:FHL_satsifaction} are invoked; otherwise $\boldsymbol{F}_{BB,r,u}$ is used as is.
\begin{proposition} \label{prop:FHL_satsifaction}
    Let $\boldsymbol{F}_{BB,r,u}^{\star}$ be a digital precoder that is compliant with both (\ref{eq:pow_constraint}) and (\ref{eq:FLr_constraint}), and $b_{r} \in [0,1)$ be FHL scaling factor for O-RU $r$, then
    \begin{equation} \label{eq:FLr_scaling}
        \boldsymbol{F}_{BB,r,u}^{\star} = b_{r} \boldsymbol{F}_{BB,r,u}, \forall u \in \setK{r}.
    \end{equation}
    Note that $\sum_{u\in\setK{r}} \rate{r}{u}$ is an increasing function over the domain of $b_r$; hence, the optimal $b_{r}$ satisfying (\ref{eq:FHL_relation2}) with equality can be obtained by line search techniques. The scaling (\ref{eq:FLr_scaling}) is iteratively repeated for all O-RUs until all FHL constraints are satisfied.
\end{proposition}
\begin{proof}
    The core part of this proof is to demonstrate that $\sum_{u\in\setK{r}} \rate{r}{u}$ is an increasing function over the domain of $b_r$, which is necessary for adopting the line search technique to ensure satisfaction of the FHL constraint. To this end, we denote the first two terms and the last term of $\boldsymbol{C}_{r,u}$ by a positive definite matrix $\boldsymbol{Q}_{r,u}$ and a positive semidefinite (PSD) matrix $\boldsymbol{O}_{r,u}$, respectively.    
    Under the scaling operation in (\ref{eq:FLr_scaling}) the rate $\rate{r}{u}$ can be re-expressed as
    \begin{multline}
        \rate{r}{u} = \log_{2} \Big| \boldsymbol{I} + \boldsymbol{F}_{BB,r,u}^{H} \boldsymbol{F}_{RF,r}^{H} \boldsymbol{H}_{r,u}^{H} \boldsymbol{W}_{r,u}^{H} \tilde{\boldsymbol{C}}_{r,u}^{-1} \\ 
        \times \boldsymbol{W}_{r,u} \boldsymbol{H}_{r,u} \boldsymbol{F}_{RF,r} \boldsymbol{F}_{BB,r,u}  \Big|,
    \end{multline}
    where $\tilde{\boldsymbol{C}}_{r,u} = \frac{1}{b_{r}^{2}}\boldsymbol{Q}_{r,u} + \boldsymbol{O}_{r,u}$ is positive definite. Notably, $\tilde{\boldsymbol{C}}_{r,u}$ decreases monotonically in the Loewner order as $b_{r}$ increases. Since matrix inversion reverses the Loewner order, it follows that the PSD matrix $\boldsymbol{N}_{r,u} (b_{r}) = \boldsymbol{F}_{BB,r,u}^{H} \boldsymbol{F}_{RF,r}^{H} \boldsymbol{H}_{r,u}^{H} \boldsymbol{W}_{r,u}^{H} \tilde{\boldsymbol{C}}_{r,u}^{-1} \boldsymbol{W}_{r,u} \boldsymbol{H}_{r,u} \boldsymbol{F}_{RF,r} \boldsymbol{F}_{BB,r,u}$ increases monotonically in the Loewner order with $b_{r}$. Next, let $ \left\{ \kappa_{r,u}^{i} (b_{r}) \geq 0 \right\}_{i=1}^{\Pi_{r,u}}$ denotes the eigenvalues of $\boldsymbol{N}_{r,u} (b_{r})$, which are used to further recast $\rate{r}{u}$ to
    \begin{equation} \label{eq:scalar_sum_ru}
        \rate{r}{u} = \sum_{i=1}^{\Pi_{r,u}} \log_{2}{\left( 1 + \kappa_{r,u}^{i} (b_{r}) \right)}.
    \end{equation}
    Because $\boldsymbol{N}_{r,u} (b_{r})$ increases monotonically with $b_{r}$ in the Loewner sense, it follows that its eigenvalues, and consequently, the rate $\rate{r}{u}$ strictly increase with $b_{r}$. Finally, recall that the sum of strictly increasing functions always results in a strictly increasing function. This property and (\ref{eq:scalar_sum_ru}) are used to assert that
    \begin{equation}
        \sum_{u\in\setK{r}} \rate{r}{u} = \sum_{u\in\setK{r}} \sum_{i=1}^{\Pi_{r,u}} \log_{2}{\left( 1 + \kappa_{r,u}^{i} (b_{r}) \right)}
    \end{equation}
    strictly increases with $b_{r}$.
\end{proof}
Appendix \ref{appendix:why_FLr_scaling} provides a simple example demonstrating the rationale for using the proposed scaling operation (\ref{eq:FLr_scaling}) to satisfy the FHL constraint.

\subsection{Optimization of $\boldsymbol{F}_{FR,r}$} \label{sect:optimize_FRF_r}
For optimization of $\boldsymbol{F}_{FR,r}$, we only activate constraint (\ref{eq:phase_r_constraint}) because (\ref{eq:pow_constraint}) is taken care of during the optimization of $\boldsymbol{F}_{BB,r,i}$. Next, we revisit (\ref{eq:convenient_obj_fn}) and drop the first two terms as they are constant w.r.t $\boldsymbol{F}_{FR,r}$. Then, we perform the following decompositions: $\bar{\boldsymbol{U}}_{r,u} \bar{\boldsymbol{U}}_{r,u}^{H} \overset{(a_{1})}{=} \evd\left( \boldsymbol{U}_{r,u} \right)$, $\boldsymbol{\Gamma}_{r,u} \boldsymbol{\Gamma}_{r,u}^{H} \overset{(a_{2})}{=} \evd\left( \sum_{\myFrac{i \in \setK{r}}{i \neq u}} \boldsymbol{F}_{BB,r,i} \boldsymbol{F}_{BB,r,i}^{H} \right)$, and $ \boldsymbol{\Delta}_{r} \boldsymbol{\Delta}_{r}^{H} \overset{(a_{3})}{=} \evd\left( \sum_{i\in\setK{r}} \boldsymbol{F}_{BB,r,i} \boldsymbol{F}_{BB,r,i}^{H} \right)$, through which we can represent (\ref{eq:convenient_obj_fn}) as
\begin{multline} \label{eq:fr_Frob_norm}
    f_{r} = \sum_{u \in \setK{r}} \left\| \bar{\boldsymbol{U}}_{r,u}^{H} \left(\boldsymbol{I}  - \boldsymbol{V}_{r,u}^{H} \boldsymbol{W}_{r,u} \boldsymbol{H}_{r,u} \boldsymbol{F}_{RF,r} \boldsymbol{F}_{BB,r,u} \right) \right\|_{F}^{2} \\
    + \sum_{u \in \setK{r}} \left \|  \bar{\boldsymbol{U}}_{r,u}^{H} \boldsymbol{V}_{r,u}^{H} \boldsymbol{W}_{r,u} \boldsymbol{H}_{r,u} \boldsymbol{F}_{RF,r} \boldsymbol{\Gamma}_{r,u} \right \|_{F}^{2} \\
    + \sum_{j \neq r}^{R} \sum_{u \in \setK{j}} \left\| \bar{\boldsymbol{U}}_{j,u}^{H} \boldsymbol{V}_{j,u}^{H} \boldsymbol{W}_{j,u} \boldsymbol{H}_{r,u} \boldsymbol{F}_{RF,r} \boldsymbol{\Delta}_{r} \right \|_{F}^{2},
\end{multline}
where $(a_{1})$, $(a_{2})$, and $(a_{3})$ hold upon left multiplication of the square root of the respective diagonal matrices by the corresponding right eigenvector matrices of the EVD operator. By letting $ \boldsymbol{E}_{1,r,u} = \boldsymbol{F}_{BB,r,u}^{T} \otimes \bar{\boldsymbol{U}}_{r,u}^{H} \boldsymbol{V}_{r,u}^{H} \boldsymbol{W}_{r,u} \boldsymbol{H}_{r,u}$, $ \boldsymbol{E}_{2,r,u} = \boldsymbol{\Gamma}_{r,u}^{T} \otimes \bar{\boldsymbol{U}}_{r,u}^{H} \boldsymbol{V}_{r,u}^{H} \boldsymbol{W}_{r,u} \boldsymbol{H}_{r,u}$, $ \boldsymbol{G}_{r,j,u} = \boldsymbol{\Delta}_{r}^{T} \otimes \bar{\boldsymbol{U}}_{j,u}^{H} \boldsymbol{V}_{j,u}^{H} \boldsymbol{W}_{j,u} \boldsymbol{H}_{r,u}$, $\boldsymbol{u}_{r,u} = \vectrz \left( \bar{\boldsymbol{U}}_{r,u}^{H} \right)$, and $\boldsymbol{f}_{RF,r} = \vectrz\left( \boldsymbol{F}_{RF,r} \right)$, (\ref{eq:fr_Frob_norm}) can further simplify to
\begin{subequations}
   \begin{align}
       f_{r} &= \sum_{u \in \setK{r}} \left\| \boldsymbol{u}_{r,u} - \boldsymbol{E}_{1,r,u} \boldsymbol{f}_{RF,r} \right\|^{2} + \sum_{u \in \setK{r}} \left \| \boldsymbol{E}_{2,r,u} \boldsymbol{f}_{RF,r}  \right \|^{2} \nonumber \\
       &+ \sum_{j \neq r}^{R} \sum_{u \in \setK{j}} \left\| \boldsymbol{G}_{r,j,u} \boldsymbol{f}_{RF,r} \right \|^{2} \\
       &= \sum_{u \in \setK{r}} \boldsymbol{u}_{r,u}^{H} \boldsymbol{u}_{r,u} - 2\Re\left( \boldsymbol{f}_{RF,r}^{H} \sum_{u \in \setK{r}} \boldsymbol{E}_{1,r,u}^{H} \boldsymbol{u}_{r,u} \right) \nonumber \\ 
       &+  \boldsymbol{f}_{RF,r}^{H} \sum_{u \in \setK{r}} \left( \boldsymbol{E}_{1,r,u}^{H} \boldsymbol{E}_{1,r,u} + \boldsymbol{E}_{2,r,u}^{H} \boldsymbol{E}_{2,r,u}\right) \boldsymbol{f}_{RF,r} \nonumber \\
       &+ \boldsymbol{f}_{RF,r}^{H} \left( \sum_{j \neq r}^{R} \sum_{u \in \setK{j}} \boldsymbol{G}_{r,j,u}^{H} \boldsymbol{G}_{r,j,u} \right) \boldsymbol{f}_{RF,r} , \label{eq:fr_obj_fn1}
   \end{align}
\end{subequations}
with the new subproblem for optimization of $\boldsymbol{F}_{RF,r}$ being 
\begin{subequations}
    \begin{align}
        \mathcal{P}(3): \min _{ \boldsymbol{f}_{RF,r} } \quad & f_{r} \\ 
        \mathrm{s.t.} \quad & \left| \left[ \boldsymbol{f}_{RF,r} \right]_{n} \right|^2 = 1, \forall_{n},
    \end{align}
\end{subequations}
whose solution can be obtained by GD techniques. As the name suggests, these techniques normally require gradient computation of the objective function, where, for the case of $f_{r}$, it is given by
\begin{multline} \label{eq:fr_grad1}
    \frac{\partial}{\partial \boldsymbol{f}_{RF,r}} f_{r} = 2\sum_{u \in \setK{r}} \left( \boldsymbol{E}_{1,r,u}^{H} \boldsymbol{E}_{1,r,u} +  \boldsymbol{E}_{2,r,u}^{H} \boldsymbol{E}_{2,r,u} \right)  \boldsymbol{f}_{RF,r}  \\
    + 2 \sum_{j \neq r}^{R} \sum_{u \in \setK{j}} \boldsymbol{G}_{r,j,u}^{H} \boldsymbol{G}_{r,j,u} \boldsymbol{f}_{RF,r} - 2\sum_{u \in \setK{r}} \boldsymbol{E}_{1,r,u}^{H} \boldsymbol{u}_{r,u}.
\end{multline}
Note that, in the course of applying the GD technique, the iterative computation of the objective function (\ref{eq:fr_obj_fn1}) and the gradient (\ref{eq:fr_grad1}) can result in high complexity, particularly because of the multiple Kronecker products and EVDs. For example, the floating point operations (FLOPs) \cite{Raphel, Golub} of the dominant operations (EVDs, Kronecker products, and matrix products $\boldsymbol{E}_{1,r,u}^{H} \boldsymbol{E}_{1,r,u}$, $\boldsymbol{E}_{2,r,u}^{H} \boldsymbol{E}_{2,r,u}$, and $\boldsymbol{G}_{r,j,u}^{H} \boldsymbol{G}_{r,j,u}$), for the computation of (\ref{eq:fr_grad1}) are given as $\mathcal{C}_{\text{legacy}} = \mathcal{O} \big( \sum_{r=1}^{R} \sum_{u \in \setK{r}} \Pi_{r,u}^{3} + \sum_{r=1}^{R} \sum_{u \in \setK{r}} M_{t}^{3} + \sum_{r=1}^{R} M_{t}^{3} + \sum_{r=1}^{R} \sum_{u \in \setK{r}} \Pi_{r,u}^{2} M_{t} N_{t} + R\sum_{r=1}^{R} \sum_{u \in \setK{r}} \Pi_{r,u} M_{t}^{2} N_{t} + \sum_{r=1}^{R} \sum_{u \in \setK{r}} ( \Pi_{r,u} M_{t} N_{t} )^{2} + R \sum_{r=1}^{R} \sum_{u \in \setK{r}} \Pi_{r,u} M_{t}^{3} N_{t}^{2} \big)$. As shown later, this complexity is high. However, it can fortunately be reduced.

To achieve that, we deploy the mixed-product property of the Kronecker product, i.e., $\left( \boldsymbol{A} \otimes \boldsymbol{B} \right) \left( \boldsymbol{C} \otimes \boldsymbol{D} \right) = \left( \boldsymbol{A} \boldsymbol{C} \right) \otimes \left( \boldsymbol{B} \boldsymbol{D} \right)$ to simplify the matrix products and their representation as follows:
\begin{subequations}
    \begin{align}
       \!\!\! \boldsymbol{E}_{1,r,u}^{H} \boldsymbol{E}_{1,r,u} &= \left( \boldsymbol{F}_{BB,r,u}^{T} \otimes \bar{\boldsymbol{U}}_{r,u}^{H} \boldsymbol{V}_{r,u}^{H} \boldsymbol{W}_{r,u} \boldsymbol{H}_{r,u} \right)^{H} \nonumber \\
        &\times \left( \boldsymbol{F}_{BB,r,u}^{T} \otimes \bar{\boldsymbol{U}}_{r,u}^{H} \boldsymbol{V}_{r,u}^{H} \boldsymbol{W}_{r,u} \boldsymbol{H}_{r,u} \right) \\
        &=\left( \boldsymbol{F}_{BB,r,u} \boldsymbol{F}_{BB,r,u}^{H} \right)^{*} \nonumber \\
        &\otimes \left( \boldsymbol{H}_{r,u}^{H} \boldsymbol{W}_{r,u}^{H} \boldsymbol{V}_{r,u} \boldsymbol{U}_{r,u} \boldsymbol{V}_{r,u}^{H} \boldsymbol{W}_{r,u} \boldsymbol{H}_{r,u} \right). \label{eq:E1ruE1ru}
    \end{align}
\end{subequations}
\begin{subequations}
    \begin{align}
       \!\!\! \boldsymbol{E}_{2,r,u}^{H} \boldsymbol{E}_{2,r,u} &= \left( \boldsymbol{\Gamma}_{r,u}^{T} \otimes \bar{\boldsymbol{U}}_{r,u}^{H} \boldsymbol{V}_{r,u}^{H} \boldsymbol{W}_{r,u} \boldsymbol{H}_{r,u} \right)^{H} \nonumber \\
        &\times \left( \boldsymbol{\Gamma}_{r,u}^{T} \otimes \bar{\boldsymbol{U}}_{r,u}^{H} \boldsymbol{V}_{r,u}^{H} \boldsymbol{W}_{r,u} \boldsymbol{H}_{r,u} \right) \\
        &=\left( \sum_{\myFrac{i \in \setK{r}}{i \neq u}} \boldsymbol{F}_{BB,r,i} \boldsymbol{F}_{BB,r,i}^{H} \right)^{*} \nonumber \\
        &\otimes \left( \boldsymbol{H}_{r,u}^{H} \boldsymbol{W}_{r,u}^{H} \boldsymbol{V}_{r,u} \boldsymbol{U}_{r,u} \boldsymbol{V}_{r,u}^{H} \boldsymbol{W}_{r,u} \boldsymbol{H}_{r,u} \right). \label{eq:E2ruE2ru}
    \end{align}
\end{subequations}
\begin{subequations}
    \begin{align}
       \!\!\! \boldsymbol{G}_{r,j,u}^{H} \boldsymbol{G}_{r,j,u} &= \left( \boldsymbol{\Delta}_{r}^{T} \otimes \bar{\boldsymbol{U}}_{j,u}^{H} \boldsymbol{V}_{j,u}^{H} \boldsymbol{W}_{j,u} \boldsymbol{H}_{r,u} \right)^{H} \nonumber \\
        &\times \left( \boldsymbol{\Delta}_{r}^{T} \otimes \bar{\boldsymbol{U}}_{j,u}^{H} \boldsymbol{V}_{j,u}^{H} \boldsymbol{W}_{j,u} \boldsymbol{H}_{r,u} \right) \\
        &=\left( \sum_{i \in \setK{r}} \boldsymbol{F}_{BB,r,i} \boldsymbol{F}_{BB,r,i}^{H} \right)^{*} \nonumber \\
        &\otimes \left( \boldsymbol{H}_{r,u}^{H} \boldsymbol{W}_{j,u}^{H} \boldsymbol{V}_{j,u} \boldsymbol{U}_{j,u} \boldsymbol{V}_{j,u}^{H} \boldsymbol{W}_{j,u} \boldsymbol{H}_{r,u} \right). \label{eq:GjruGjru}
    \end{align}
\end{subequations}
For starters, observe that (\ref{eq:E1ruE1ru}), (\ref{eq:E2ruE2ru}), and (\ref{eq:GjruGjru}), collectively dissolves the need for the EVD of $\boldsymbol{U}_{r,u}$, $\sum_{\myFrac{i \in \setK{r}}{i \neq u}} \boldsymbol{F}_{BB,r,i} \boldsymbol{F}_{BB,r,i}^{H}$, and $\sum_{i \in \setK{r}} \boldsymbol{F}_{BB,r,i} \boldsymbol{F}_{BB,r,i}^{H}$.
Nonetheless, the terms $\boldsymbol{u}_{r,u}^{H} \boldsymbol{u}_{r,u}$ and $\boldsymbol{E}_{1,r,u}^{H} \boldsymbol{u}_{r,u}$ in (\ref{eq:fr_obj_fn1}) and (\ref{eq:fr_grad1}), respectively, still depend on the EVD of $\boldsymbol{U}_{r,u}$. To completely drop this dependence, we express these terms as follows:
\begin{equation}
    \boldsymbol{u}_{r,u}^{H} \boldsymbol{u}_{r,u} =  \Tr\left(\bar{\boldsymbol{U}}_{r,u} \bar{\boldsymbol{U}}_{r,u}^{H} \right) = \Tr\left( \boldsymbol{U}_{r,u} \right).
\end{equation}
\begin{equation} \label{eq:E1uru}
    \boldsymbol{E}_{1,r,u}^{H} \boldsymbol{u}_{r,u} \overset{(b)}{=} \vectrz \left( \boldsymbol{H}_{r,u}^{H} \boldsymbol{W}_{r,u}^{H} \boldsymbol{V}_{r,u} \boldsymbol{U}_{r,u} \boldsymbol{F}_{BB,r,u}^{H}\right),
\end{equation}
where $(b)$ holds from the ``\textit{vec trick}'', $\left( \boldsymbol{B}^{T} \otimes \boldsymbol{A} \right) \vectrz\left( \boldsymbol{X} \right) = \vectrz\left( \boldsymbol{A} \boldsymbol{X} \boldsymbol{B} \right)$. Furthermore, the revelation in (\ref{eq:E1ruE1ru}) and (\ref{eq:E2ruE2ru}) shows that the sum term $\boldsymbol{E}_{1,r,u}^{H} \boldsymbol{E}_{1,r,u} +  \boldsymbol{E}_{2,r,u}^{H} \boldsymbol{E}_{2,r,u}$ in (\ref{eq:fr_obj_fn1}) and (\ref{eq:fr_grad1}) can be further simplified to
\begin{align}
    & \boldsymbol{E}_{1,r,u}^{H} \boldsymbol{E}_{1,r,u} +  \boldsymbol{E}_{2,r,u}^{H} \boldsymbol{E}_{2,r,u} \nonumber \\
    &= \left( \sum_{i \in \setK{r}} \boldsymbol{F}_{BB,r,i} \boldsymbol{F}_{BB,r,i}^{H} \right)^{*} \nonumber \\
    &\otimes \left( \boldsymbol{H}_{r,u}^{H} \boldsymbol{W}_{r,u}^{H} \boldsymbol{V}_{r,u} \boldsymbol{U}_{r,u} \boldsymbol{V}_{r,u}^{H} \boldsymbol{W}_{r,u} \boldsymbol{H}_{r,u} \right) \label{eq:sum_E1ru_E2ru}.
\end{align}
Now, putting together the simplifications (\ref{eq:GjruGjru}) -- (\ref{eq:sum_E1ru_E2ru}), the objective function (\ref{eq:fr_obj_fn1}) and its derivative (\ref{eq:fr_grad1}) take the final forms
\begin{multline} \label{eq:fr_obj_fn2}
   \!\!\!\!\! f_{r} = \boldsymbol{f}_{RF,r}^{H} \left( \sum_{j = 1 }^{R} \sum_{u \in \setK{j}} \boldsymbol{G}_{r,j,u}^{H} \boldsymbol{G}_{r,j,u} \right) \boldsymbol{f}_{RF,r} - 2 \Re\left( \boldsymbol{f}_{RF,r}^{H} \boldsymbol{s}_{r} \right)\! \\
   + \sum_{u \in \setK{r}} \Tr\left( \boldsymbol{U}_{r,u} \right)
\end{multline}
and
\begin{equation} \label{eq:fr_grad2}
    \frac{\partial}{\partial \boldsymbol{f}_{RF,r}} f_{r} = 2 \sum_{j = 1}^{R} \sum_{u \in \setK{j}} \boldsymbol{G}_{r,j,u}^{H} \boldsymbol{G}_{r,j,u} \boldsymbol{f}_{RF,r} - 2\boldsymbol{s}_{r},
\end{equation}
respectively, which are free of EVDs and perform a single Kronecker product per O-RU. The last claim in the previous statement is true due to the fact that the terms $\sum_{j = 1}^{R} \sum_{u \in \setK{j}} \boldsymbol{G}_{r,j,u}^{H} \boldsymbol{G}_{r,j,u} = \left( \sum_{i \in \setK{r}} \boldsymbol{F}_{BB,r,i} \boldsymbol{F}_{BB,r,i}^{H} \right)^{*} \otimes \big( \sum_{j = 1}^{R} \sum_{u \in \setK{j}} \boldsymbol{H}_{r,u}^{H} \boldsymbol{W}_{r,u}^{H} \boldsymbol{V}_{j,u} \boldsymbol{U}_{j,u} \boldsymbol{V}_{j,u}^{H} \boldsymbol{W}_{j,u} \boldsymbol{H}_{r,u} \big)$ and $ \boldsymbol{s}_{r} =  \vectrz \big( \sum_{u \in \setK{r}} \boldsymbol{H}_{r,u}^{H} \boldsymbol{W}_{r,u}^{H} \boldsymbol{V}_{r,u} \boldsymbol{U}_{r,u} \boldsymbol{F}_{BB,r,u}^{H}\big)$, whose dominant complexities are $ \mathcal{C}_{G_{r}} = \mathcal{O} \big( M_{t}^{2} N_{t}^{2} + \sum_{j=1}^{R} \sum_{u \in \setK{j}} N_{t}^{2} \Pi_{j,u} \big)$ and $ \mathcal{C}_{s_{r}} = \mathcal{O} \big( \sum_{u \in \setK{r}} M_{t} \Pi_{r,u}^{2} + N_{t} \Pi_{r,u} \big( N_{r} + M_{t} \big) \big)$, respectively, appear in both (\ref{eq:fr_obj_fn2}) and (\ref{eq:fr_grad2}), hence they can be computed once and reused.
Given the expressions in (\ref{eq:fr_obj_fn2}) and (\ref{eq:fr_grad2}), a natural question arises: \textit{why begin with EVDs?} The use of EVDs at the outset of the solution serves to expose the hidden properties of the problem that are not readily apparent from a direct examination of (\ref{eq:convenient_obj_fn}). As demonstrated in this subsection, these properties are subsequently leveraged to simplify the solution and reduce its overall complexity, including the eventual exclusion of EVDs.

\begin{figure}
    \centering
    \subfloat[FLOPs vs. the number of transmit antennas ($N_{t}$).]{
        \label{Fig:FLOPs_vs_Nt}
        \includegraphics[width = \figW\textwidth]{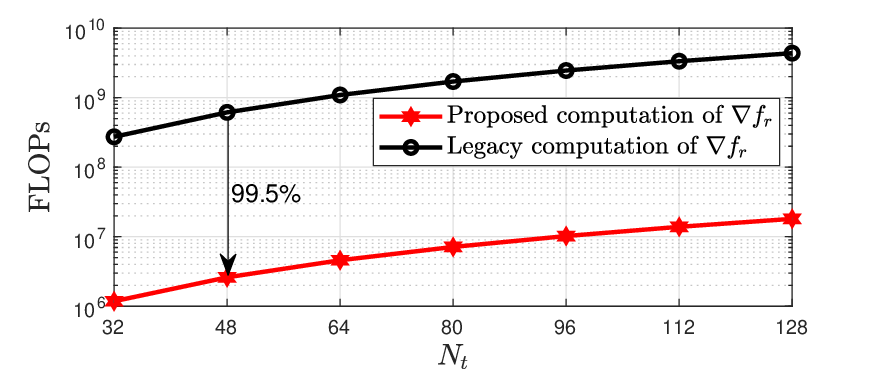}
    }
    
    \subfloat[FLOPs vs. the number of streams ($\Pi_{r,u}$).]{
        \label{Fig:FLOPs_vs_Streams}
        \includegraphics[width = \figW\textwidth]{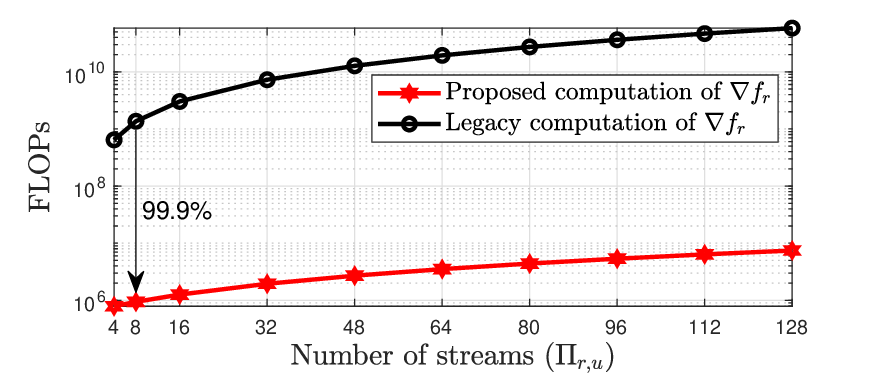}
    }
    
    \caption{FLOPs comparison between the proposed and legacy computation of $\frac{d}{d \boldsymbol{f}_{RF,r}} f_{r} \equiv \nabla f_{r}$.}
    \label{Fig:Prop_vs_Legacy_FLOPs}
    % \vspace{-5mm}
\end{figure}

In summary, to solve $\mathcal{P}(3)$ by GD techniques, a low-complexity objective (\ref{eq:fr_obj_fn2}) and its respective gradient (\ref{eq:fr_grad2}), are adopted. To demonstrate the relevance of the additional steps (\ref{eq:E1ruE1ru}) -- (\ref{eq:sum_E1ru_E2ru}), we compare the complexity of computing the derivative of $f_{r}$ by using (\ref{eq:fr_grad2}) and (\ref{eq:fr_grad1}) in \cref{Fig:Prop_vs_Legacy_FLOPs}. As observed in Figs. \ref{Fig:Prop_vs_Legacy_FLOPs} \subref{Fig:FLOPs_vs_Nt} and \subref{Fig:FLOPs_vs_Streams}, the proposed computation of $\frac{d}{d \boldsymbol{f}_{RF,r}} f_{r}$ by (\ref{eq:fr_grad2}) consistently reduces the complexity by over 99\% for various numbers of transmit antennas and number of streams, respectively.

\subsection{Optimization of $\boldsymbol{W}_{FR,u}$} 
\label{sect:optimize_WRF_u}
The solution for $\boldsymbol{W}_{FR,u}$ is very similar to that of $\boldsymbol{F}_{FR,r}$. This is due to the high similarity in their mode of operation and hardware constraints. Following this similarity, the solution used for the latter is adopted for the former. In the following, we briefly describe how the solution for $\boldsymbol{W}_{FR,u}$ can be obtained with reduced complexity. By letting $ \bar{\boldsymbol{C}}_{r,u} = \sigma_{u}^{2} \boldsymbol{I} + \sum_{l \neq r}^{R} \boldsymbol{H}_{l,u} \boldsymbol{F}_{RF,l} \sum_{i \in \setK{l}} \boldsymbol{F}_{BB,l,i} \boldsymbol{F}_{BB,l,i}^{H}  \boldsymbol{F}_{RF,l}^{H} \boldsymbol{H}_{l,u}^{H} + \boldsymbol{H}_{r,u} \boldsymbol{F}_{RF,r} \sum_{ \myFrac{j \in \setK{r}}{j \neq u} } \boldsymbol{F}_{BB,r,j} \boldsymbol{F}_{BB,r,j}^{H} \boldsymbol{F}_{RF,r}^{H} \boldsymbol{H}_{r,u}^{H} $, whose EVD is given as $\boldsymbol{\Psi}_{r,u} \boldsymbol{\Psi}_{r,u}^{H} = \evd\left( \bar{\boldsymbol{C}}_{r,u} \right)$, allows the recast of (\ref{eq:convenient_obj_fn}) to
\begin{align}
    & f_{u} \nonumber \\
    & = \sum_{r \in \setR{u}} \left\| \bar{\boldsymbol{U}}_{r,u}^{H} \left( \boldsymbol{I}  - \boldsymbol{V}_{r,u}^{H} \boldsymbol{W}_{BB,u,r} \boldsymbol{W}_{RF,u} \boldsymbol{T}_{r,u} \right) \right\|_{F}^{2} \nonumber \\
    & + \sum_{r \in \setR{u}} \left \|  \bar{\boldsymbol{U}}_{r,u}^{H} \boldsymbol{V}_{r,u}^{H} \boldsymbol{W}_{BB,u,r} \boldsymbol{W}_{RF,u} \boldsymbol{\Psi}_{r,u} \right \|_{F}^{2} \nonumber \\
    & = \sum_{r \in \setR{u}} \left\| \boldsymbol{u}_{r,u} -  \left( \boldsymbol{T}_{r,u}^{T} \otimes  \bar{\boldsymbol{U}}_{r,u}^{H} \boldsymbol{V}_{r,u}^{H} \boldsymbol{W}_{BB,u,r} \right) \boldsymbol{w}_{RF,u} \right\|^{2} \nonumber \\
    & + \sum_{r \in \setR{u}} \left \| \left( \boldsymbol{\Psi}_{r,u}^{T} \otimes \bar{\boldsymbol{U}}_{r,u}^{H} \boldsymbol{V}_{r,u}^{H} \boldsymbol{W}_{BB,u} \right) \boldsymbol{w}_{RF,u} \right \|^{2} \nonumber \\
    & = \sum_{r \in \setR{u}} \left\| \boldsymbol{u}_{r,u} -  \boldsymbol{E}_{3,r,u} \boldsymbol{w}_{RF,u} \right\|^{2} + \sum_{r \in \setR{u}} \left\| \boldsymbol{E}_{4,r,u} \boldsymbol{w}_{RF,u} \right\|^{2} \nonumber \\
    & = \sum_{r \in \setR{u}} \boldsymbol{u}_{r,u}^{H} \boldsymbol{u}_{r,u} - 2\Re\left( \boldsymbol{w}_{RF,u}^{H} \sum_{r \in \setR{u}} \boldsymbol{E}_{3,r,u}^{H} \boldsymbol{u}_{r,u} \right) \nonumber \\
    &+ \boldsymbol{w}_{RF,u}^{H} \sum_{r \in \setR{u}} \left( \boldsymbol{E}_{3,r,u}^{H} \boldsymbol{E}_{3,r,u} + \boldsymbol{E}_{4,r,u}^{H} \boldsymbol{E}_{4,r,u} \right) \boldsymbol{w}_{RF,u} \label{eq:fu_Frob_norm},
\end{align}
where $ \boldsymbol{E}_{3,r,u} = \boldsymbol{T}_{r,u}^{T} \otimes  \bar{\boldsymbol{U}}_{r,u}^{H} \boldsymbol{V}_{r,u}^{H} \boldsymbol{W}_{BB,u,r}$, $ \boldsymbol{E}_{4,r,u} = \boldsymbol{\Psi}_{r,u}^{T} \otimes \bar{\boldsymbol{U}}_{r,u}^{H} \boldsymbol{V}_{r,u}^{H} \boldsymbol{W}_{BB,u,r} $, $\boldsymbol{w}_{RF,u} = \vectrz\left( \boldsymbol{W}_{RF,u} \right)$, and $\boldsymbol{T}_{r,u} = \boldsymbol{H}_{r,u} \boldsymbol{F}_{RF,r} \boldsymbol{F}_{BB,r,u}$ is temporarily used to compress the notation.
As demonstrated previously, the requirement of EVDs and multiple Kronecker products for each O-RU–user pair renders the evaluation of (\ref{eq:fu_Frob_norm}) and its derivative computationally expensive. To reduce this complexity, the mixed-product property of the Kronecker product is adopted once again. Specifically, by following arguments similar to those in (\ref{eq:E1ruE1ru}), (\ref{eq:E2ruE2ru}), (\ref{eq:E1uru}), and (\ref{eq:sum_E1ru_E2ru}), (\ref{eq:fu_Frob_norm}) reduces to
\begin{equation} \label{eq:fu_obj}
    f_{u} \overset{(c)}{=} \boldsymbol{w}_{RF,u}^{H} \boldsymbol{J}_{u} \boldsymbol{w}_{RF,u} - 2 \Re\left( \boldsymbol{w}_{RF,u}^{H}   \boldsymbol{s}_{u}  \right ), 
\end{equation}
with its derivative w.r.t $\boldsymbol{w}_{RF,u}$ being
\begin{equation} \label{eq:fu_grad}
    \frac{\partial}{\partial \boldsymbol{w}_{RF,u}} f_{u} =   2 \left( \boldsymbol{J}_{u} \boldsymbol{w}_{RF,u} - \boldsymbol{s}_{u} \right),
\end{equation}
where
\begin{multline}
    \boldsymbol{J}_{u} = \Bigg( \sum_{l =1}^{R} \boldsymbol{H}_{l,u} \boldsymbol{F}_{RF,l} \sum_{i \in \setK{l}} \boldsymbol{F}_{BB,l,i} \boldsymbol{F}_{BB,l,i}^{H}  \boldsymbol{F}_{RF,l}^{H} \boldsymbol{H}_{l,u}^{H} \\
    +\sigma_{u}^{2} \boldsymbol{I} \Bigg)^{*} \otimes \left( \sum_{r \in \setR{u}} \boldsymbol{W}_{BB,u,r}^{H} \boldsymbol{V}_{r,u} \boldsymbol{U}_{r,u} \boldsymbol{V}_{r,u}^{H} \boldsymbol{W}_{BB,u,r}\right)
\end{multline}
$\boldsymbol{s}_{u} = \vectrz\left( \sum_{r \in \setR{u}} \boldsymbol{W}_{BB,u,r}^{H}\boldsymbol{V}_{r,u} \boldsymbol{U}_{r,u} \boldsymbol{F}_{BB,r,u}^{H} \boldsymbol{F}_{RF,r}^{H} \boldsymbol{H}_{r,u}^{H} \right)$, and $(c)$ is due to the drop of the first term of (\ref{eq:fu_Frob_norm}).
Consequently, the resulting subproblem for the optimization of $\boldsymbol{w}_{RF,u}$, given as
\begin{subequations}
    \begin{align}
        \mathcal{P}(4): \min _{ \boldsymbol{w}_{RF,u} } \quad & f_{u} \\ 
        \mathrm{s.t.} \quad & \left| \left[ \boldsymbol{w}_{RF,u} \right]_{n} \right|^2 = 1, \forall_{n},
    \end{align}
\end{subequations}
can be solved by GD techniques. Similar to the previous subsection, because the terms $ \boldsymbol{J}_{u} $ and $ \boldsymbol{s}_{u}$, whose dominant complexities are $ \mathcal{C}_{J_{u}} = \mathcal{O} \left( N_{r}^{2} M_{r}^{2} \right)$ and $\mathcal{C}_{s_{u}} = \mathcal{O} \big( \sum_{r \in \setR{u}} \Pi_{r,u} \big( M_{r}\Pi_{r,u} + M_{t}\Pi_{r,u} + N_{t} N_{r} + M_{t} N_{t} + M_{r} N_{r}\big) \big)$, respectively, appear in both (\ref{eq:fu_obj}) and (\ref{eq:fu_grad}), their values can be computed once and reused. Also note that the Kronecker product is computed once per user rather than for each O-RU-user combination.

\subsection{Optimization of $\boldsymbol{W}_{BB,u,r}$} \label{sect:optimize_WBB_u_r}
Observe that, when all other variables are fixed and the design of $\boldsymbol{W}_{BB,u,r}$ is considered, the objective function of $\mathcal{P}(1)$ reduces to (\ref{eq:Rate_ru}). Accordingly, we reformulate (\ref{eq:Rate_ru}) to explicitly expose the optimizable variable $\boldsymbol{W}_{BB,u,r}$ as
\begin{multline} \label{eq:Rate_ru_WBB_ru}
    \rate{r}{u} = \log_{2} \Big| \boldsymbol{I} + \big(\boldsymbol{W}_{BB,u,r} \hat{\boldsymbol{C}}_{r,u}  \boldsymbol{W}_{BB,u,r}^{H} \big)^{-1} \boldsymbol{W}_{BB,u,r} \\
    \times \boldsymbol{L}_{r,u} \boldsymbol{L}_{r,u}^{H} \boldsymbol{W}_{BB,u,r}^{H} \Big|,
\end{multline}
where $\hat{\boldsymbol{C}}_{r,u} = \boldsymbol{W}_{FR,u} \bar{\boldsymbol{C}}_{r,u} \boldsymbol{W}_{FR,u}^{H}$ and $\boldsymbol{L}_{r,u} = \boldsymbol{W}_{RF,u} \boldsymbol{H}_{r,u} \boldsymbol{F}_{RF,r} \boldsymbol{F}_{BB,r,u}$. Finding $\boldsymbol{W}_{BB,u,r}$ that maximizes (\ref{eq:Rate_ru_WBB_ru}) in its current form, however, remains challenging. To simplify the optimization, we let $\hat{\boldsymbol{W}}_{BB,u,r} = \boldsymbol{W}_{BB,u,r} \hat{\boldsymbol{C}}_{r,u}^{1/2}$ denote a proxy optimizable variable for $\boldsymbol{W}_{BB,u,r}$, which simplifies (\ref{eq:Rate_ru_WBB_ru}) to
\begin{multline}
    \hat{\mathcal{R}}_{r,u} = \log_{2} \Big| \boldsymbol{I} + \big(\hat{\boldsymbol{W}}_{BB,u,r}\hat{\boldsymbol{W}}_{BB,u,r}^{H} \big)^{-1} \hat{\boldsymbol{W}}_{BB,u,r} \hat{\boldsymbol{C}}_{r,u}^{-1/2} \\
    \times \boldsymbol{L}_{r,u} \boldsymbol{L}_{r,u}^{H} \hat{\boldsymbol{C}}_{r,u}^{-1/2} \hat{\boldsymbol{W}}_{BB,u,r}^{H} \Big|,
\end{multline}
with its corresponding problem being
\begin{equation}
    \mathcal{P}(5): \max_{ \hat{\boldsymbol{W}}_{BB,u,r} } \quad  \hat{\mathcal{R}}_{r,u}.
\end{equation}
To solve $\mathcal{P}(5)$, we adopt a simple strategy of designing $\hat{\boldsymbol{W}}_{BB,u,r}$ such that the received streams $\Pi_{r,u}$ are mutually parallel. Specifically, let the columns of $\boldsymbol{\Omega} \in \cdim{M_{r}}{M_{r}}$ denote the right orthonormal eigenvectors associated with the ordered eigenvalues $\gamma_{r,u,1} > \gamma_{r,u,2} > \cdots > \gamma_{r,u,M_{r}}$ of the PSD matrix $\hat{\boldsymbol{C}}_{r,u}^{-1/2} \boldsymbol{L}_{r,u} \boldsymbol{L}_{r,u}^{H} \hat{\boldsymbol{C}}_{r,u}^{-1/2}$. Then the solution to $\mathcal{P}(5)$ is obtained by $\big[ \hat{\boldsymbol{W}}_{BB,u,r} \big]_{i,:} = \big[ \boldsymbol{\Omega}^{H} \big]_{i,:}, i \in 1,\cdots,\Pi_{r,u}$, where $\big[ \boldsymbol{X} \big]_{i,:}$ denotes the $i$-th row of $\boldsymbol{X}$. Finally, the digital combiner is recovered from the proxy variable as
\begin{equation} \label{eq:WBB_ur_soln}
    \boldsymbol{W}_{BB,u,r} = \hat{\boldsymbol{W}}_{BB,u,r} \hat{\boldsymbol{C}}_{r,u}^{-1/2}.
\end{equation}
Using (\ref{eq:WBB_ur_soln}), it can be verified that the objective function (\ref{eq:Rate_ru_WBB_ru}) reduces to
\begin{equation} \label{eq:Rate_ru_gamma_ru}
    \rate{r}{u} =  \sum_{i=1}^{\Pi_{r,u}} \log_{2} \left(1 + \gamma_{r,u,i} \right),
\end{equation}
with $\gamma_{r,u,i}$ equivalently denoting the signal-to-interference plus noise ratio (SINR) of the $i$-th stream of user $u$ from O-RU $r$.

\subsection{The overall algorithm}
\begin{algorithm}
    \DontPrintSemicolon
    \SetAlgoLined
    \SetKwInOut{Input}{Input}
    \SetKwInOut{Output}{Output}
    \Input{ $ \left\{ \sigma_{u}^{2} \right\}_{u=1}^{K} $, $ \left\{ P_{r} \right\}_{r=1}^{R} $, $ \left\{ FL_{r} \right\}_{r=1}^{R} $, and $ \boldsymbol{H}_{r,u}, \forall r,u$. }
    
    Initialization: Randomly generate $ \boldsymbol{F}_{BB,r,u} $, $ \boldsymbol{W}_{BB,u,r}, \forall r,u $, $\left\{ \boldsymbol{F}_{RF,r} \right\}_{r=1}^{R}$, and $\left\{ \boldsymbol{W}_{RF,u} \right\}_{u = 1}^{K}$ to their feasible values.\;
    $I_{o} = 0$.\;
    \Repeat{$ \mathrm{(\ref{eq:convenient_obj_fn})} ~ \mathrm{converges~or~} I_{o} \geq I_{max}$}{
        Obtain $\boldsymbol{W}_{BB,u,r}, \forall u,r$ according to (\ref{eq:WBB_ur_soln}). \;
        $I_{W_{RF}} = 0$.\;
		\Repeat{$ (\ref{eq:fu_obj}) \mathrm{~converges~or~} I_{W_{RF}} \geq I_{max} $}{ \label{step:update_WRF_start}
            Update $ \boldsymbol{V}_{r,u} $ and $ \boldsymbol{U}_{r,u}, \forall r,u$. \;
            Solve $ \mathcal{P}(4) $ to obtain $\boldsymbol{W}_{RF,u}, \forall u$. \;
            $I_{W_{RF}} \leftarrow I_{W_{RF}} + 1$.\;
		} \label{step:update_WRF_end}
        $I_{F_{RF}} = 0$.\;
        \Repeat{$ (\ref{eq:fr_obj_fn2}) \mathrm{~converges~or~} I_{F_{RF}} \geq I_{max} $}{ \label{step:update_FRF_start}
            Update $ \boldsymbol{V}_{r,u} $ and $ \boldsymbol{U}_{r,u}, \forall r,u$. \;
            Solve $ \mathcal{P}(3) $ to obtain $\boldsymbol{F}_{RF,r}, \forall r$. \;
            $I_{F_{RF}} \leftarrow I_{F_{RF}} + 1$.\;
		} \label{step:update_FRF_end}
        $I_{F_{BB}} = 0$.\;
        \Repeat{$ I_{F_{BB}} \geq I_{max} \mathrm{~or~}$ $ \sum_{r = 1}^{R} \sum_{u \in \setK{r}} \boldsymbol{U}_{r,u} \mathrm{~converges}$ }{ \label{step:update_FBB_rru_start}
            Update $ \boldsymbol{V}_{r,u} $ and $ \boldsymbol{U}_{r,u}, \forall r,u$. \;
            Obtain $ \boldsymbol{F}_{BB,r,u}, \forall r,u $ according to (\ref{eq:FBB_ru_soln}). \;
            $I_{F_{BB}} \leftarrow I_{F_{BB}} + 1$.\;
		} \label{step:update_FBB_rru_end}

        $I_{o} \leftarrow I_{o} + 1$.\;
    }
     Scale $ \boldsymbol{F}_{BB,r,u}$ to obtain $ \boldsymbol{F}_{BB,r,u}^{\star}, \forall r,u $ \label{step:FBB_ru_scaling}.
     
    \Output{$ \boldsymbol{F}_{BB,r,u}^{\star}, \forall r, u $, $\left\{ \boldsymbol{W}_{BB,u}, \boldsymbol{W}_{RF,u} \right\}_{u = 1}^{K}$, and $\left\{ \boldsymbol{F}_{RF,r} \right\}_{r=1}^{R}$.}

    \caption{Summary of the proposed solution.}
    \label{alg:soln_P1}
\end{algorithm}
\cref{alg:soln_P1} summarizes these procedures. The algorithm takes as inputs the channel matrices, FHLs capacities, O-RUs transmit power, and users' noise variance. Using these inputs, it randomly initializes\footnote{Other initialization techniques can be adopted such as zero forcing for $\boldsymbol{F}_{BB,r,u}$ and codebook-based vectors for analog precoders/combiners. However, in our tests, none of these approaches outperformed random initialization.} both digital and analog precoders/combiners and iteratively updates them until convergence or until the maximum number of iterations, $I_{max}$, is reached.
As shown in step \ref{step:FBB_ru_scaling}, after the algorithm converges, the digital precoders for each O-RU--user pair are scaled to ensure that the power-compliant digital precoders also satisfy the FHL constraint. Owing to the nature of the problems $\mathcal{P}(3)$ and $\mathcal{P}(4)$, techniques such as manifold optimization (MO) \cite{AbsMahSep2008}, which is adopted in this work, or projected gradient descent (PGD) \cite{7849224}, with Armijo line search for step size selection are guaranteed to converge to a local optimum. The MO or PGD procedure is terminated when the maximum number of iterations is reached or when the difference between the objective-function values at two successive iterations falls below a $10^{-5}$ tolerance.
Similarly, because the solution of the digital precoder follows a block coordinate descent update, involving both the precoder and its auxiliary variable, convergence is guaranteed according to \cite{Solodov98}.
Although \cref{alg:soln_P1} has a nested structure, note that each inner loop can be considered an individual block. Therefore, the preceding argument applies to establish the convergence of the overall algorithm.

\subsection{Complexity analysis}
We use FLOPs to analyze the computational complexity of the proposed algorithm. 
The complexity for the update of the digital combiners is dominated by the inverse operation, of which total FLOPS is given by $\mathcal{C}_{W_{BB}} =  \mathcal{O} \big( \sum_{r=1}^R \sum_{u \in \setK{r}} M_{r}^{3} \big) $.
Next, the complexity for the update of $\boldsymbol{V}_{r,u}$'s and $\boldsymbol{U}_{r,u}$'s are dominated by their inverse operations, of which computational complexities are given by $ \mathcal{C}_{V_{r,u}} =  \mathcal{O}\big( \Pi_{r,u}^{3} \big)$ and $\mathcal{C}_{U_{r,u}} = \mathcal{O} \big( \Pi_{r,u}^{3} \big)$, respectively. However, the update of $\boldsymbol{V}_{r,u}$'s must be preceded by the computation of $\boldsymbol{C}_{r,u}$, of which FLOPs is given by $\mathcal{C}_{C_{r,u}} = \mathcal{O} \big( N_{r}N_{t}M_{t} + N_{r}M_{t}^{2} + N_{r}^{2}M_{t} + KM_{t}^2\Pi_{r,u} \big)$.
Therefore, the total complexity for the update of $\boldsymbol{W}_{RF,u}$'s and $\boldsymbol{F}_{RF,r}$'s is given by $ \mathcal{C}_{W_{RF}} =  I_{W_{RF}} \sum_{u=1}^{K} \sum_{r \in \setR{u}} \big( \mathcal{C}_{C_{r,u}} + \mathcal{C}_{V_{r,u}} + \mathcal{C}_{U_{r,u}} \big) + I_{W_{RF}} K \big( \mathcal{C}_{J_{u}} + \mathcal{C}_{s_{u}} + I_{GD} \big( M_{r}N_{r} \big)^2 \big)$ and $ \mathcal{C}_{F_{RF}} = I_{F_{RF}} R \big( \mathcal{C}_{G_{r}} + \mathcal{C}_{s_{r}} + I_{GD} \big( M_{t}N_{t} \big)^2 \big) + I_{F_{RF}} \sum_{r=1}^{R} \sum_{u \in \setK{r}} \big( \mathcal{C}_{C_{r,u}} + \mathcal{C}_{V_{r,u}} + \mathcal{C}_{U_{r,u}} \big) $, respectively, where $I_{GD}$, $I_{W_{RF}}$, and $I_{F_{RF}}$ is the number of iterations for the GD, the update of $\boldsymbol{W}_{RF,u}$, and $\boldsymbol{F}_{RF,r}$, respectively.

Similarly, the complexity for the update of digital precoders is given by $ \mathcal{C}_{F_{BB}} = I_{F_{BB}} R \mathcal{C}_{X_{r}} + I_{F_{BB}} \sum_{r=1}^{R} \sum_{u \in \setK{r}} \big( \mathcal{C}_{C_{r,u}} + \mathcal{C}_{V_{r,u}} + \mathcal{C}_{U_{r,u}} \big)$, where $I_{F_{BB}}$ is the number of iterations for the update of $\boldsymbol{F}_{BB,r,u}$'s and $\mathcal{C}_{X_{r}} = \mathcal{O} \big( M_{t}^{3} \big)$ is the number of FLOPs for the dominant inverse operation in (\ref{eq:FBB_ru_soln}).
Consequently, the total complexity of \cref{alg:soln_P1} is given by $\mathcal{C}_{\text{Alg. 1}} = I_{o} \big( \mathcal{C}_{W_{BB}} + \mathcal{C}_{W_{RF}} + \mathcal{C}_{F_{RF}} + \mathcal{C}_{F_{BB}} \big) $, where $I_{o}$ is the number of iterations for the outer loop.

\section{Simulation Results} \label{Sect:Simulations}

\subsection{Parameter setup}
Unless otherwise stated, we adopt the following network parameters. A BS disaggregated into $R=4$ O-RUs, each with a transmit power of $P_{r} = \Power$ dBm and $M_{t} = 8$ RF chains fully connected to $N_{t} = 24$ transmit antennas via $N_{t}M_{t}$ phase shifters, serves a set of $K=4$ users. Each user is located at a distance of $D_{r,u} = \Du$ m from the BS and experiences a noise variance of $\sigma_{u}^{2} = -87$ dBm. Each O-RU is connected to the O-DU through an independent FHL with a throughput of $FL_{r} = 8$ Gbps. To ensure almost lossless recovery of the quantized precoders and combiners, we use a large number of bits, i.e., $q_{a} = q_{d} = 22$ \cite{10478417}, to represent each quantization level.
The transceiver architecture of the O-RUs is mirrored at the user side, where each user is equipped with $N_{r} = 12$ receive antennas connected to $M_{r} = 8$ RF chains via $N_{r}M_{r}$ fully connected phase shifters. The antenna gain at each O-RU is 24.5 dBi, while the antenna gain at each user is 15 dBi \cite{11165760}. Each O-RU transmits $\Pi_{r,u} = 2$ streams to each user. The system is allocated a bandwidth of $BW = 100$ MHz centered at 28 GHz carrier frequency.
Similar to \cite{11165760}, we adopt a millimeter-wave channel model with 4 propagation paths and a slot duration of $t_{slot} = 1$ ms to characterize the small-scale fading, whereas the large-scale fading is modeled as $PL(D_{r,u}) [\text{dB}] = 128.1 +37.6\log_{10}\big( D_{r,u}/1000 \big)$. Finally, we set $R_{use} = 1$, $R_{code} = 1$, assume $O_{r} = 0$ bits, and use (\ref{eq:P1_obj_fn}) to generate all SE-related results.

\subsection{Benchmark schemes}
For benchmarking purposes, the following schemes are considered:
\begin{enumerate}
    \item \textbf{non-GoB}: The analog precoders and combiners are designed using a non-GoB approach, while the digital precoder is obtained by solving $\mathcal{P}(2)$ via CVX \cite{11165760}.
    The dominant complexity for this scheme is given by $\mathcal{C}_{\text{non-GoB}} = \mathcal{O} \big( \sum_{r=1}^{R} \sum_{u \in \setK{r}}\big( N_{t} N_{r} M_{t} + M_{t}M_{r}^{2} \big) \big) + \mathcal{O} \big( \sum_{r=1}^{R} \sum_{u \in \setK{r}} N_{t} N_{r}^{2} \big) + \mathcal{O} \big( \sum_{r=1}^{R} \sum_{u \in \setK{r}}\big( N_{t} N_{r} M_{r} + N_{t}M_{r}^{2} \big) \big) + I_{\text{CVX}} \mathcal{O} \big( (M_{t}\Pi_{r,u})^{3.5}\log(1/\epsilon) + \sum_{r=1}^{R} \sum_{u \in \setK{r}}\big( \mathcal{C}_{C_{r,u}} + \mathcal{C}_{V_{r,u}} + \mathcal{C}_{U_{r,u}} \big) \big)$, where $I_{\text{CVX}}$ is the number of times CVX is invoked and $\epsilon = 10^{-2}$ is the numerical accuracy of the solver.
    \item \textbf{Weighted MMSE and OMP (WMMSE–OMP)}: This scheme is a variant of \cite[Alg. 4]{8048606}, extended in this work to support the multi-transmitter scenario under consideration.
    Extending from \cite[Table 1]{8048606}, the complexity of this scheme is given by $\mathcal{C}_{\text{OMP}} = \mathcal{O} \big( I_{OMP} \sum_{r=1}^{R} \sum_{u \in \setK{r}} N^{3} N_{RF} \big)$, where $N = \max\{N_{t}, N_{r} \}$, $N_{RF} = \max\{M_{t}, M_{r} \}$, and $I_{OMP}$ is the number of iterations needed by this scheme.
    \item \textbf{Successive interference rejection MMSE (SIR–MMSE)}: As proposed in \cite{9747002}, the analog precoders and combiners are designed by minimizing the Frobenius norm of the corresponding effective channel, while the digital precoders and combiners are successively designed using an interference rejection strategy.
    This process incurs the following complexity $\mathcal{C}_{\text{SIR}} = I_{SIR} \mathcal{O} \big( \sum_{r=1}^{R} N_{t}^{3} + \sum_{u=1}^{K} N_{r}^{3} \big) + \mathcal{O} \big( \sum_{r=1}^{R} \sum_{u \in \setK{r}} M_{t}^{3} + N_{t}^{2}N_{r} \big)$, where $I_{SIR}$ is the number of iteration needed for convergence of the analog beamforming.
    \item \textbf{Two-stage digital precoding}: This scheme is a variant of the proposed \cref{alg:soln_P1}, except that the digital precoders at each O-RU are designed following the two-stage digital precoding approach in \cite{7160780}.
    Accordingly, we compute the dominant complexity of this scheme as follows: $\mathcal{C}_{\text{TS}} = I_{o} \mathcal{O} \big( \mathcal{C}_{W_{BB}} + \mathcal{C}_{W_{RF}} + \mathcal{C}_{F_{RF}} \big) + \mathcal{O}\big( \sum_{r=1}^{R} \sum_{u\in \setK{r}} N_{r}M_{t}N_{t} + N_{r}M_{t}\Pi_{r,u} + N_{r}M_{r}\Pi_{r,u} \big) + \mathcal{O} \big( \sum_{r=1}^{R} \daleth_{r}^{2} (M_{t} + \daleth_{r}) \big) $, where $\daleth_{r} = \sum_{u\in\setK{r}} \Pi_{r,u}$.
    \item \textbf{ZF digital precoding}: This scheme is another variant of the proposed \cref{alg:soln_P1}, where each O-RU designs its digital precoders according to a ZF criterion, i.e., by neglecting user noise and inter-O-RU interference. Specifically, by letting $ \boldsymbol{F}_{BB,r} = \left[ \boldsymbol{F}_{BB,r,i}, \cdots, \boldsymbol{F}_{BB,r,|\setK{r}|} \right] \in \cdim{Mt}{ \sum_{u \in \setK{r}} \Pi_{r,u} }, i \in \setK{r}$ to be the digital precoders of O-RU $r$ for users in $\setK{r}$, then
    \begin{equation}
        \boldsymbol{F}_{BB,r} = \sqrt{P_{r}} \left(\left\| \boldsymbol{F}_{RF,r} \hat{\boldsymbol{F}}_{BB,r}\right\|_{F}\right)^{-1} \hat{\boldsymbol{F}}_{BB,r},
    \end{equation}
    where $ \hat{\boldsymbol{F}}_{BB,r} = \boldsymbol{M}_{r}^{H} \big( \boldsymbol{M}_{r} \boldsymbol{M}_{r}^{H} \big)^{-1}$, $\boldsymbol{M}_{r} =  \boldsymbol{W}_{BB,r} \boldsymbol{W}_{RF} \boldsymbol{H}_{r} \boldsymbol{F}_{RF,r}$, $\boldsymbol{W}_{BB,r} = \text{blkdiag} \big( \boldsymbol{W}_{BB,i,r}, \cdots, \boldsymbol{W}_{BB,|\setK{r}|,r} \big)$, $\boldsymbol{W}_{RF} = \text{blkdiag} \big( \boldsymbol{W}_{RF,i}, \cdots, \boldsymbol{W}_{RF,|\setK{r}|} \big)$, and $\boldsymbol{H}_{r} = \Big[ \boldsymbol{H}_{r,i}^{T}, \cdots, \boldsymbol{H}_{r,|\setK{r}|}^{T} \Big]^{T}, \forall i \in \setK{r}$.
    Similarly, its complexity is given by $\mathcal{C}_{\text{ZF}} = I_{o} \mathcal{O} \big( \mathcal{C}_{W_{BB}} + \mathcal{C}_{W_{RF}} + \mathcal{C}_{F_{RF}} \big) + \mathcal{O}\big( \sum_{r=1}^{R} \big( N_{t}M_{t} +\daleth_{r}M_{t} + \daleth_{r} \sum_{u\in\setK{r}} M_{r} \big) \sum_{u\in\setK{r}} N_{r} \big) + \mathcal{O} \big( \sum_{r=1}^{R} \daleth_{r}^{2} (M_{t} + \daleth_{r}) \big) $.
\end{enumerate}

\subsection{Convergence results}
\begin{figure}
    \centering
    \includegraphics[width = \figW\textwidth]{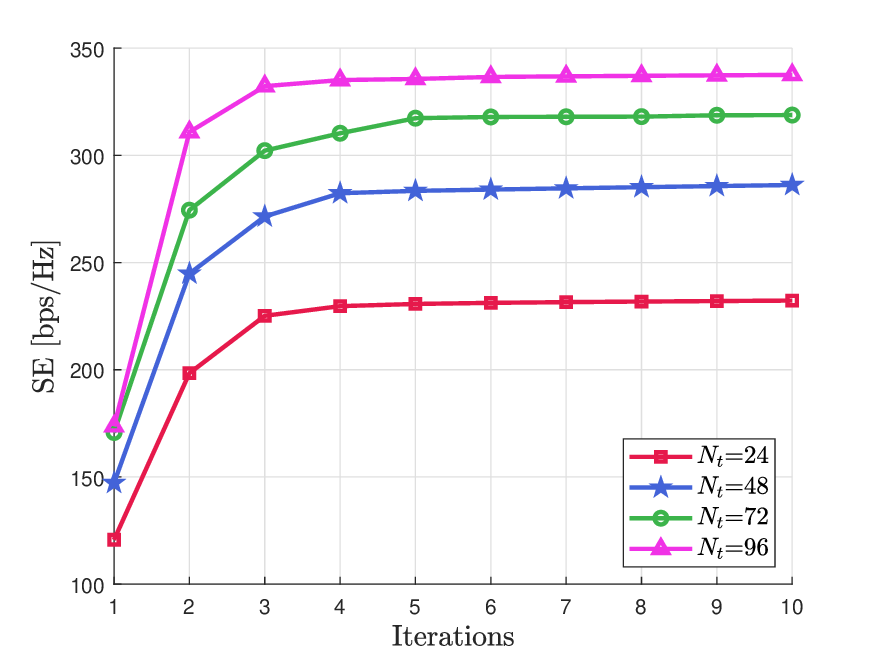}
    \caption{Convergence behavior of the proposed algorithm for various numbers of transmit antennas at each O-RU.}
    \label{Fig:SE_vs_Iters}
    % \vspace{-5mm}
\end{figure}

We evaluate the convergence behavior of the proposed algorithm for different numbers of transmit antennas and present the corresponding results for four configurations in \cref{Fig:SE_vs_Iters}. Two key observations can be drawn from \cref{Fig:SE_vs_Iters}. First, the proposed algorithm exhibits monotonic convergence across all configurations, indicating robust and stable behavior under varying system dimensions. Second, convergence is achieved rapidly, typically within five iterations--highlighting the practicality of the proposed algorithm.

\begin{figure}
    \centering
    \includegraphics[width = \figW\textwidth]{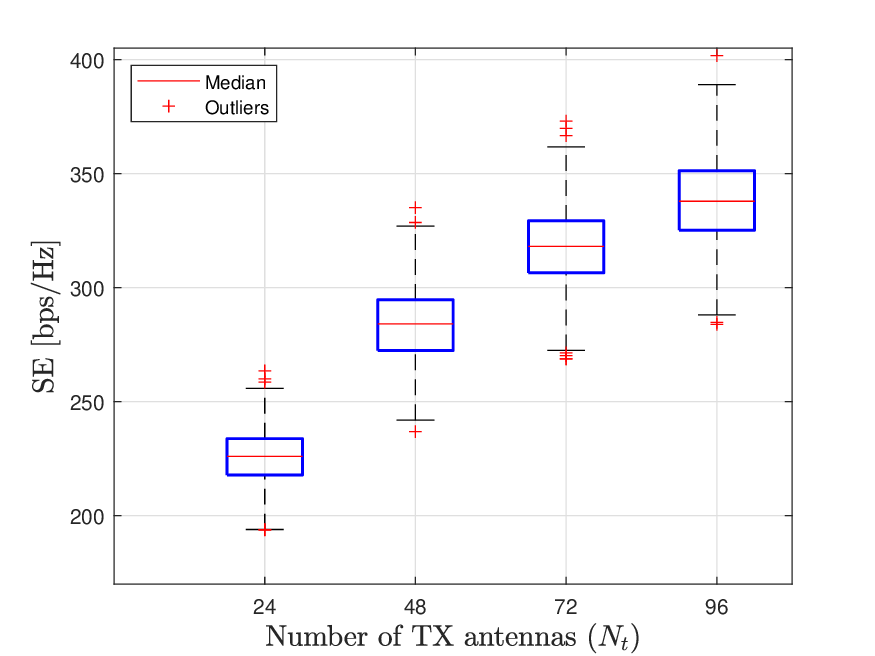}
    \caption{Sensitivity of the proposed algorithm to various random initialization points.}
    \label{Fig:SE_vs_inits}
    % \vspace{-5mm}
\end{figure}

Because the proposed algorithm relies on random initialization, we further evaluate the impact of different random initial points in \cref{Fig:SE_vs_inits}. The median performance is consistent with the average convergence values reported in \cref{Fig:SE_vs_Iters}, indicating that the proposed algorithm consistently achieves favorable solutions across different random initializations.

\subsection{Achievable rate results}
This subsection evaluates the SE performance and computational complexity of the proposed algorithm in comparison with benchmark schemes under various network configurations. To comprehensively assess the advantages of the proposed algorithm, these results are presented across multiple system settings.

\begin{figure}
    \centering
    \includegraphics[width = \figW\textwidth]{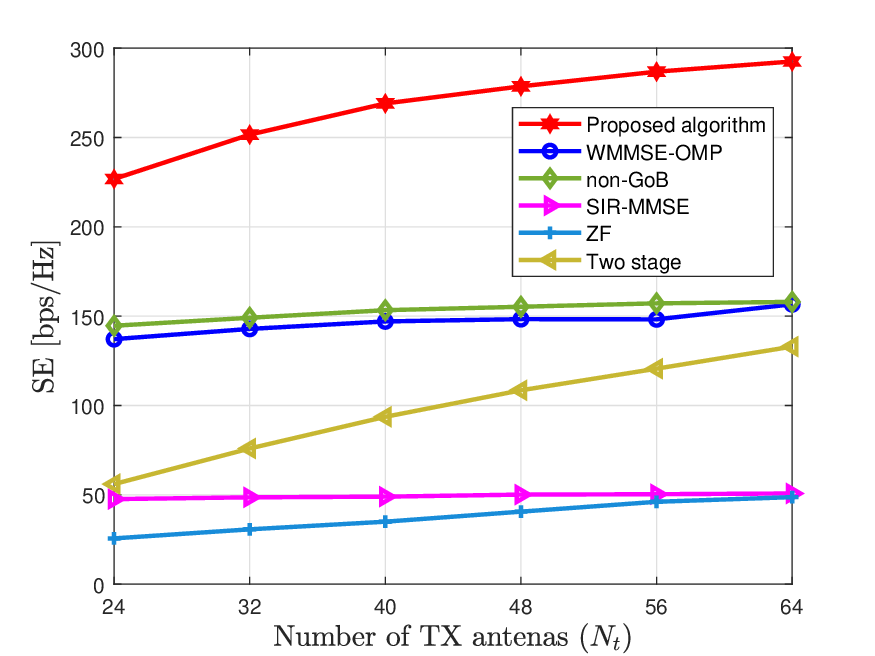}
    \caption{SE vs. the number of transmit antennas at each O-RU when $P_{r} = 30$ dBm, $\Pi_{r,u} = 2$, $FL_{r} = 8$ Gbps, $R = 4$, $K = 4$, $M_{t} = M_{r} = 8$, and $N_{r} = 12$.}
    \label{Fig:SE_vs_Nt}
    % \vspace{-5mm}
\end{figure}

First, we vary the number of transmit antennas per O-RU and report the corresponding SE performance of all schemes in \cref{Fig:SE_vs_Nt}. As expected, the SE of all schemes increases with $N_{t}$ due to the enhanced array gain afforded by larger antenna arrays. Notably, the proposed algorithm consistently achieves the highest SE among all schemes. This trend, which is also observed in the subsequent figures, highlights the superior performance of the proposed algorithm relative to the benchmark schemes.
Interestingly, the performance gap between the proposed algorithm and the best-performing benchmark scheme increases with the number of transmit antennas. As shown in \cref{Fig:SE_vs_Nt}, the performance gap between the proposed algorithm and the non-GoB scheme increases from 82 to 134 bps/Hz as the number of transmit antennas increases from 24 to 64.
Qualitatively, we note that most benchmark schemes employ suboptimal designs for either the analog or digital beamformers, resulting in inferior overall performance. For instance, the analog beamformers in the WMMSE–OMP scheme are constrained by the size and quality of predefined codebooks, whereas the proposed algorithm designs analog beamformers without such limitations, offering significantly greater design flexibility. Similarly, the analog beamformers in the SIR–MMSE and non-GoB schemes are designed using heuristic criteria: maximizing the Frobenius norm of local effective channels in the former and minimizing the Frobenius norm of constrained and unconstrained beamformers in the latter, which ultimately leads to suboptimal performance. Applying analogous reasoning to the ZF and Two-stage schemes, we observe that although they employ analog beamformers from the proposed algorithm, their digital precoder designs neglect user noise and inter-O-RU interference, substantially degrading their performance.

\begin{figure}
    \centering
    \includegraphics[width = \figW\textwidth]{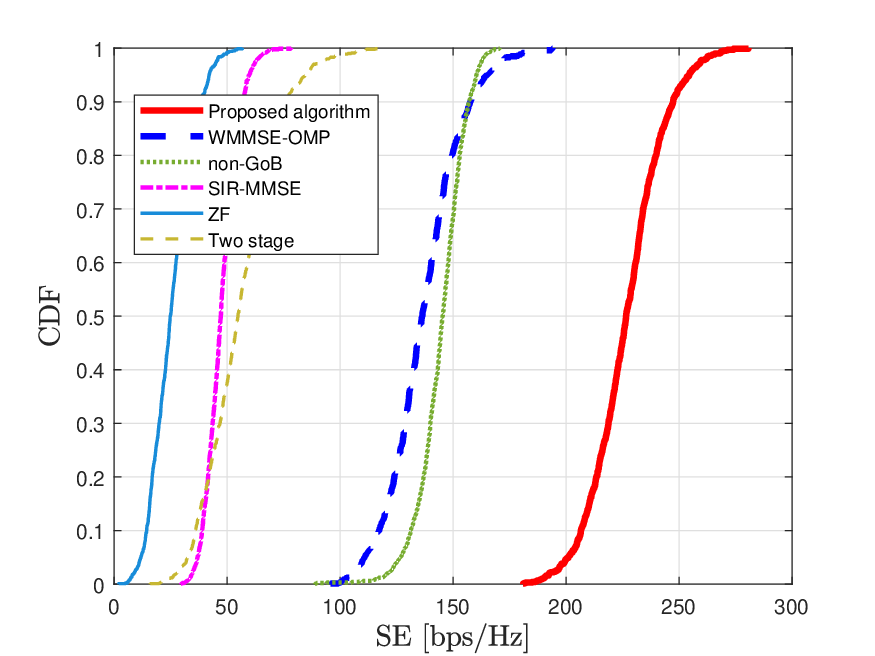}
    \caption{CDF of the network SE for $N_{t} = 24$, $P_{r} = 30$ dBm, $\Pi_{r,u} = 2$, $FL_{r} = 8$ Gbps, $R = 4$, $K = 4$, $M_{t} = M_{r} = 8$, and $N_{r} = 12$.}
    \label{Fig:cdf_vs_SE}
    % \vspace{-5mm}
\end{figure}

\cref{Fig:cdf_vs_SE} further extends this comparison by illustrating the cumulative distribution function (CDF) of the network SE for $N_{t}=24$. Consistent with previous observations, the proposed algorithm outperforms all benchmark schemes by a wide margin. Quantitatively, the proposed algorithm achieves the SE of at least 175 bps/Hz almost deterministically, whereas the non-GoB and WMMSE–OMP schemes attain the same performance for less than 1\% of the time, and the remaining benchmark schemes fail to meet this threshold. Increasing this target by only 25 bps/Hz to SE $\geq$ 200 bps/Hz, the proposed algorithm still meets this requirement approximately 95\% of the time, while none of the benchmark schemes do. These results further underscore the reliability of the proposed algorithm in consistently achieving high SE performance.

\begin{figure}
    \centering
    \includegraphics[width = \figW\textwidth]{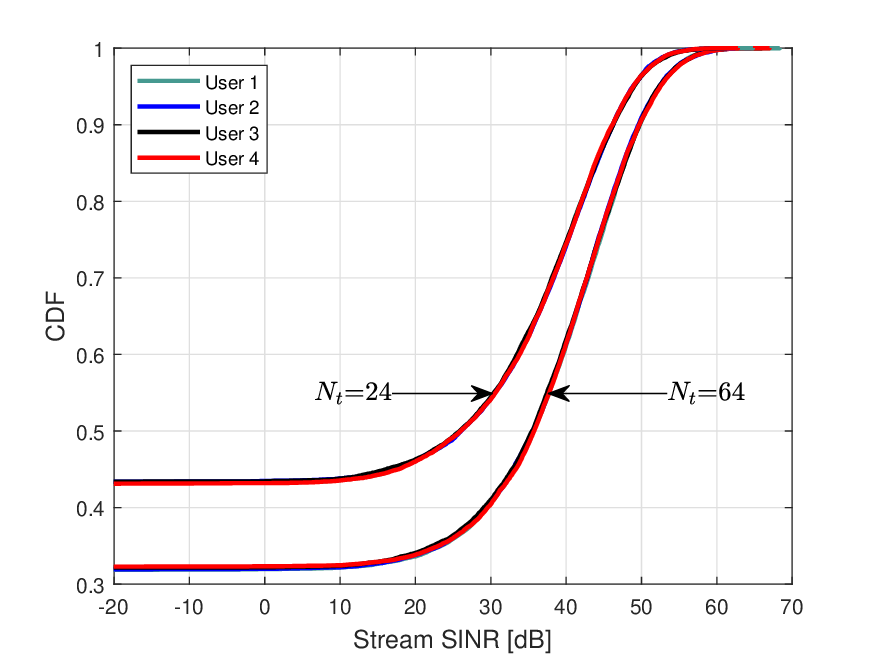}
    \caption{CDF of user stream SINR when $P_{r} = 30$ dBm, $\Pi_{r,u} = 2$, $FL_{r} = 8$ Gbps, $R = 4$, $K = 4$, $M_{t} = M_{r} = 8$, and $N_{r} = 12$.}
    \label{Fig:SINR_vs_streams}
    % \vspace{-5mm}
\end{figure}

\cref{Fig:SINR_vs_streams} unpacks the performance of the proposed algorithm reported in \cref{Fig:SE_vs_Nt} for $N_{t}$ = 24 and 64 into the SINR CDFs of each user's all $\aleph_{u}$ streams. As expected, the streams of systems with more transmit antennas achieve higher SINRs than those of systems with fewer transmit antennas. This improvement can be attributed to the increased transmit-array diversity gain and the effectiveness of the proposed algorithm in realizing it.
Moreover, as shown in \cref{Fig:SINR_vs_streams}, the SINR of a user's stream exceeds $10\log_{10}(\gamma_{r,u,i})=10$ dB in approximately 67\% and 56\% of the considered channel realizations for $N_{t}=64$ and $24$, respectively. We also observe that when a user $u$ experiences a relatively weaker channel with O-RU $r$ than the other users, its $\Pi_{r,u}$ streams tend to be allocated less transmit power. This behavior is consistent with the objective of maximizing the system-wide sum rate, under which allocating more power to streams with stronger effective channels can yield a larger increase in the aggregate achievable rate.
This observation also helps explain the lower tail of the CDFs in \cref{Fig:SINR_vs_streams}, where there is approximately a 33\% and 44\% probability that a user's stream has an SINR below 0 dB for $N_{t}=64$ and $24$, respectively.

\begin{figure}
    \centering
    \includegraphics[width = \figW\textwidth]{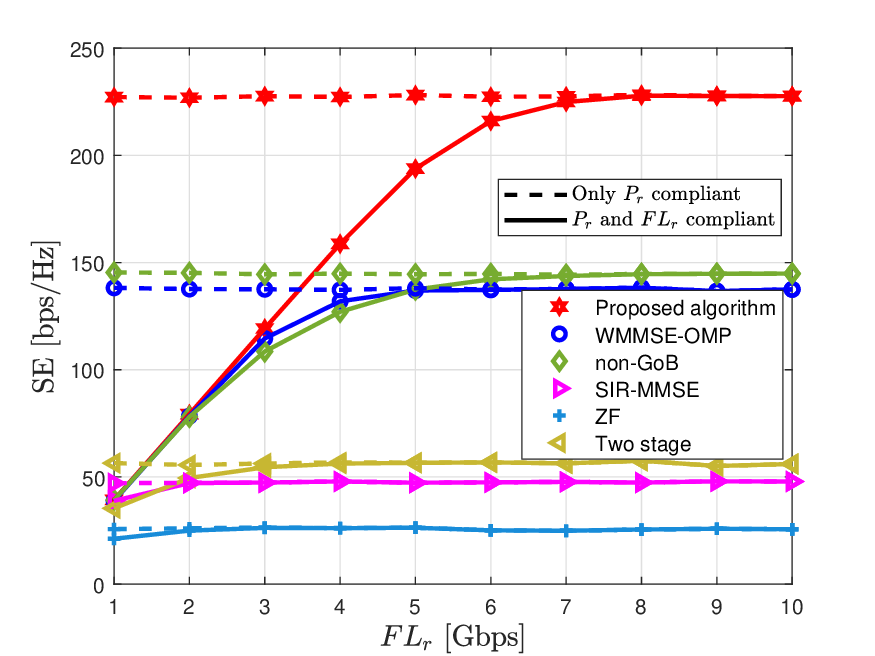}
    \caption{Analysis of the impact of FHL capacity when $N_{t} = 24$, $P_{r} = 30$ dBm, $\Pi_{r,u} = 2$, $R = 4$, $K = 4$, $M_{t} = M_{r} = 8$, and $N_{r} = 12$.}
    \label{Fig:SE_vs_FHL}
    % \vspace{-5mm}
\end{figure}

To investigate the impact of the FHL capacity constraint $FL_{r}$ and the effectiveness of the proposed scaling operation in (\ref{eq:FLr_scaling}), we present in \cref{Fig:SE_vs_FHL} the SE performance of all schemes under varying FHL capacities. When $FL_{r}$ is small, all schemes except ZF converge to similar SE levels. This behavior arises because, as the scaling operation enforces the FHL constraint by adjusting each scheme's digital precoders, it forces even more capable algorithms to underutilize their available transmit power. Consequently, under stringent FHL constraints, all schemes deliver comparable SE performance.
Effectively, the scaling operation (\ref{eq:FLr_scaling}) allows each algorithm to first maximize SE and then reduce the transmit power only when the FHL constraint is violated. This leads to a scenario where, under tight FHL constraints, all schemes operate at reduced transmit power levels, thereby improving their energy efficiency (EE). This behavior highlights an additional benefit of the proposed scaling mechanism.

Another noteworthy trend in \cref{Fig:SE_vs_FHL} is that, as $FL_{r}$ increases, the SE of all schemes gradually improves until reaching a saturation point. Beyond this point, the FHL constraint becomes inactive, allowing all schemes to fully exploit the available transmit power. Note that under these conditions, the proposed algorithm achieves the highest EE, primarily due to its superior SE performance.

\begin{figure}
    \centering
    \includegraphics[width = \figW\textwidth]{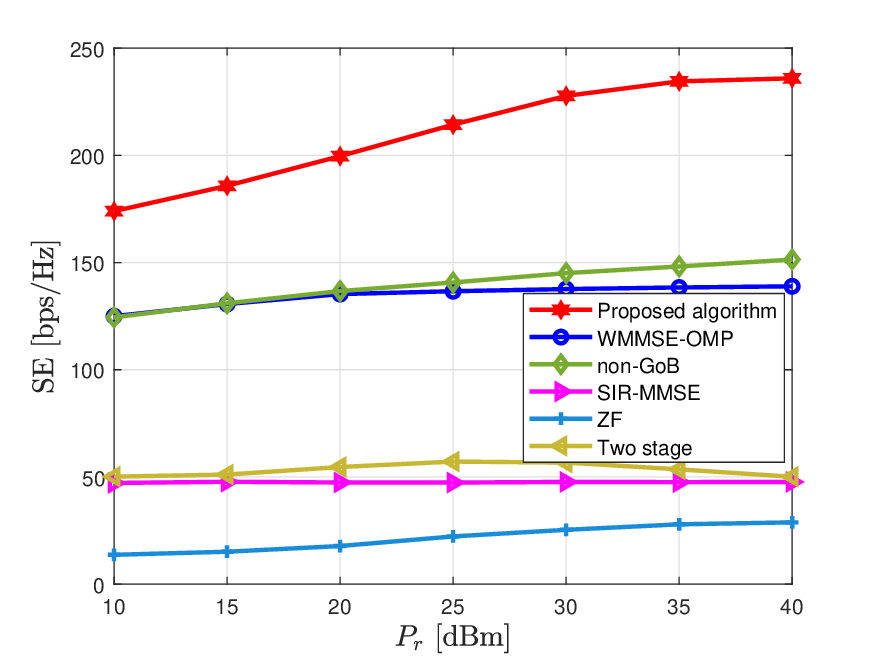}
    \caption{SE vs. O-RU transmit power ($P_{r}$) when $N_{t} = 24$, $\Pi_{r,u} = 2$, $FL_{r} = 8$ Gbps, $R = 4$, $K = 4$, $M_{t} = M_{r} = 8$, and $N_{r} = 12$.}
    \label{Fig:SE_vs_Pr}
    % \vspace{-5mm}
\end{figure}

To further examine the impact of transmit power, \cref{Fig:SE_vs_Pr} shows the SE performance response as the transmit power of each O-RU varies from 10 dBm to 40 dBm. As expected, the SE of most schemes increases with transmit power. The robustness of the proposed algorithm in jointly designing analog and digital beamformers enables it to maintain superior performance even in power-limited regimes. Moreover, the performance gap between the proposed algorithm and benchmark schemes widens with increasing transmit power, further indicating its superior energy efficiency.

\begin{figure}
    \centering
    \includegraphics[width = \figW\textwidth]{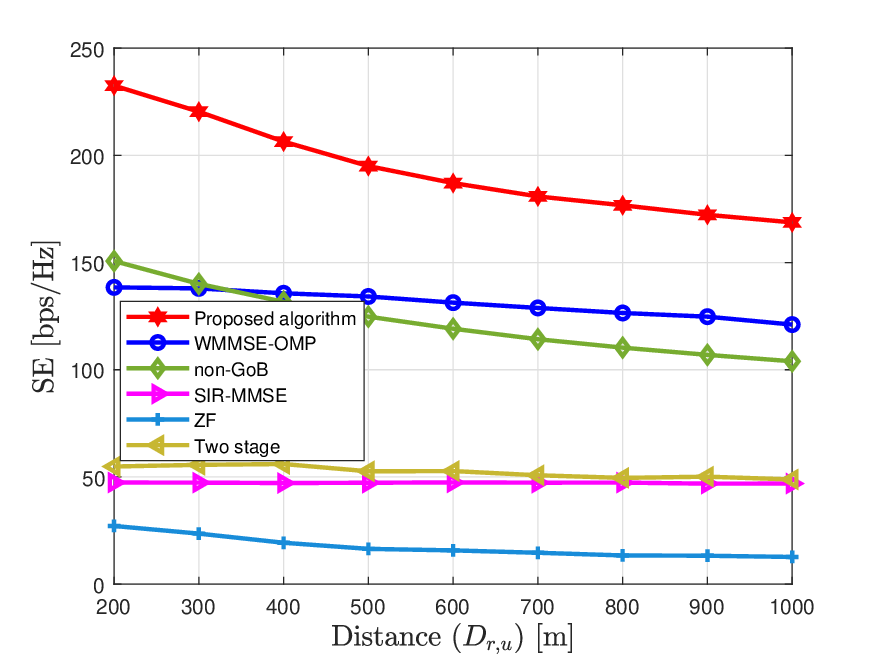}
    \caption{SE vs. O-RU-user distance ($D_{r,u}$) when $N_{t} = 24$, $P_{r} = 30$ dBm, $\Pi_{r,u} = 2$, $FL_{r} = 8$ Gbps, $R = 4$, $K = 4$, $M_{t} = M_{r} = 8$, and $N_{r} = 12$.}
    \label{Fig:SE_vs_D}
    % \vspace{-5mm}
\end{figure}

In \cref{Fig:SE_vs_D}, it is observed that the SE decreases as the distance between the BS and the users increases. This expected behavior is due to large-scale fading increasing with distance; therefore, the signal attenuates more as it travels farther, resulting in the observed reduced SE. Consistent with earlier results, the proposed algorithm achieves the highest performance across all considered BS–user distances, further demonstrating its superiority over the benchmark schemes.

\begin{figure}
    \centering
    \includegraphics[width = \figW\textwidth]{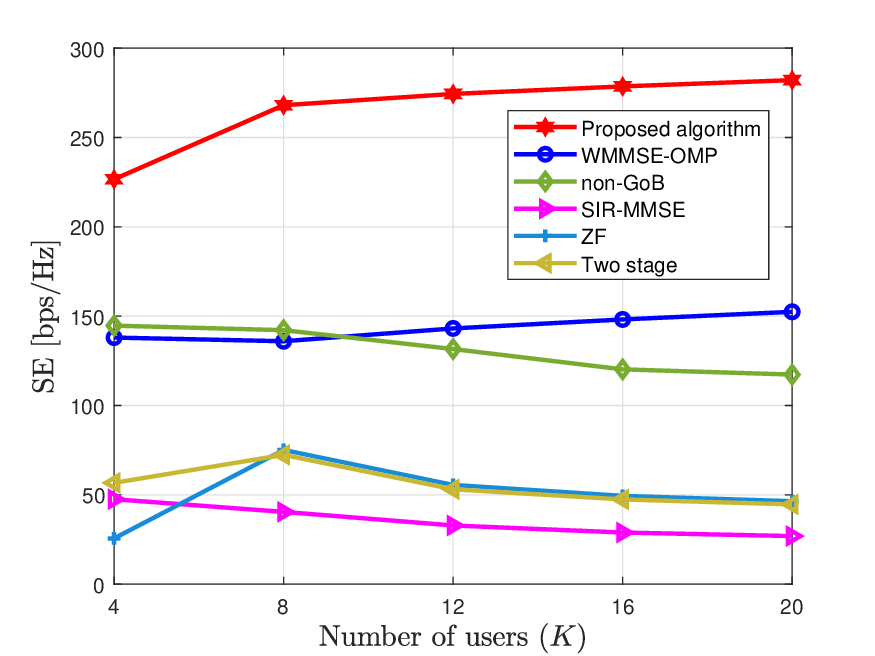}
    \caption{SE vs. number of users $K$ when $N_{t} = 24$, $P_{r} = 30$ dBm, $\Pi_{r,u} = 2$, $FL_{r} = 8$ Gbps, $R = 4$, $M_{t} = M_{r} = 8$, and $N_{r} = 12$.}
    \label{Fig:SE_vs_K}
    % \vspace{-5mm}
\end{figure}

We further evaluate the performance of the various schemes for different numbers of users in \cref{Fig:SE_vs_K}. On the one hand, most benchmark schemes exhibit performance degradation as the number of users increases. This behavior can be attributed to increased demand under fixed system resources, which degrades the overall quality of service. On the other hand, the proposed algorithm and the WMMSE-OMP scheme maintain a slight performance improvement as the number of users increases. This trend highlights their effectiveness in resource allocation and in mitigating both user noise and inter-user interference. Although both WMMSE-OMP and the proposed algorithm demonstrate an increasing performance trend, the proposed algorithm consistently achieves higher SE, underscoring its superior beamforming capability.

\subsection{Analysis and impact of channel estimation errors}
The results in the previous subsection assume that the O-DU has perfect knowledge of the channel state information (CSI). Nonetheless, practical systems must first perform channel estimation, a process that is usually error-prone, to obtain a CSI that is subsequently used to design various parameters. Because of these errors, the designed parameters may be misaligned with the true channel, leading to performance degradation. To model the impact of these errors, we let $\boldsymbol {H}_{r,u}$ denote the true channel for the estimated channel $\hat{\boldsymbol{H}}_{r,u}$ and assume that the estimation process introduces, on each element of $\hat{\boldsymbol{H}}_{r,u}$, a normalized mean square error (NMSE) of $\rho^{4}$, i.e., 
\begin{equation} \label{eq:chan_nmse}
    \frac{\mathbb{E} \Big\{ \Big| \big[ \boldsymbol{H}_{r,u} \big]_{i,j} - \big[ \hat{\boldsymbol{H}}_{r,u} \big]_{i,j} \Big|^{2} \Big\}}{\sigma_{h}^{2}} = \rho^{4}, \forall i,j,
\end{equation}
where $\sigma_{h}^{2} = \mathbb{E} \Big\{ \Big|  \big[ \hat{\boldsymbol{H}}_{r,u} \big]_{i,j} \Big|^{2} \Big\}$. To achieve (\ref{eq:chan_nmse}), the true and estimated channels are related as $\boldsymbol{H}_{r,u} = \hat{\boldsymbol{H}}_{r,u} + \rho^{2} \sqrt{\sigma_{h}^{2}} \boldsymbol{\Xi}$, where $\rho \geq 0$ denotes the magnitude of CSI estimation error and each element of $\boldsymbol{\Xi} \in \cdim{N_{r}}{N_{t}}$ is a circularly symmetric complex Gaussian random variable with zero mean and unit variance.

\begin{figure}
    \centering
    \includegraphics[width = \figW\textwidth]{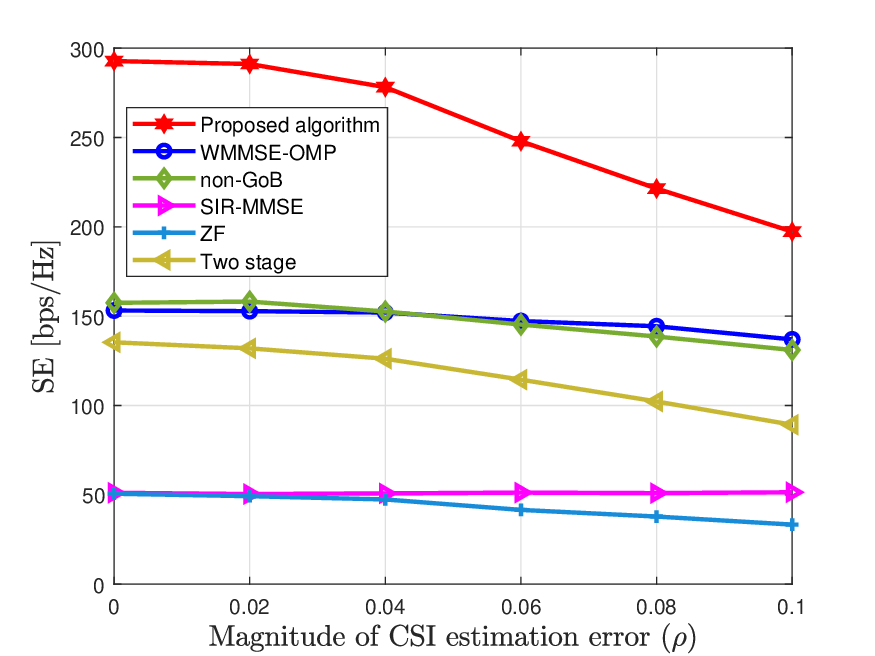}
    \caption{SE vs. magnitude of CSI estimation error when $N_{t} = 64$, $P_{r} = 30$ dBm, $\Pi_{r,u} = 2$, $FL_{r} = 8$ Gbps, $R = K = 4$, $M_{t} = M_{r} = 8$, and $N_{r} = 12$.}
    \label{Fig:SE_vs_CSI_error_Nt64}
    % \vspace{-5mm}
\end{figure}

Using this model, we plot the performance response of all the schemes for different magnitudes of CSI estimation error in \cref{Fig:SE_vs_CSI_error_Nt64}. As expected, the performance of most schemes degrades as the estimation error increases. However, this performance degradation can be mitigated by designing the system parameters while accounting for the NMSE of the CSI estimation process. Future work will consider enhancing the proposed algorithm's robustness to CSI estimation errors through such a design approach.
Despite the observed performance degradation, the proposed algorithm consistently outperforms all benchmark schemes over the entire considered range of $\rho$.

\subsection{Complexity analysis}
\begin{figure}
    \centering
    \includegraphics[width = \figW\textwidth]{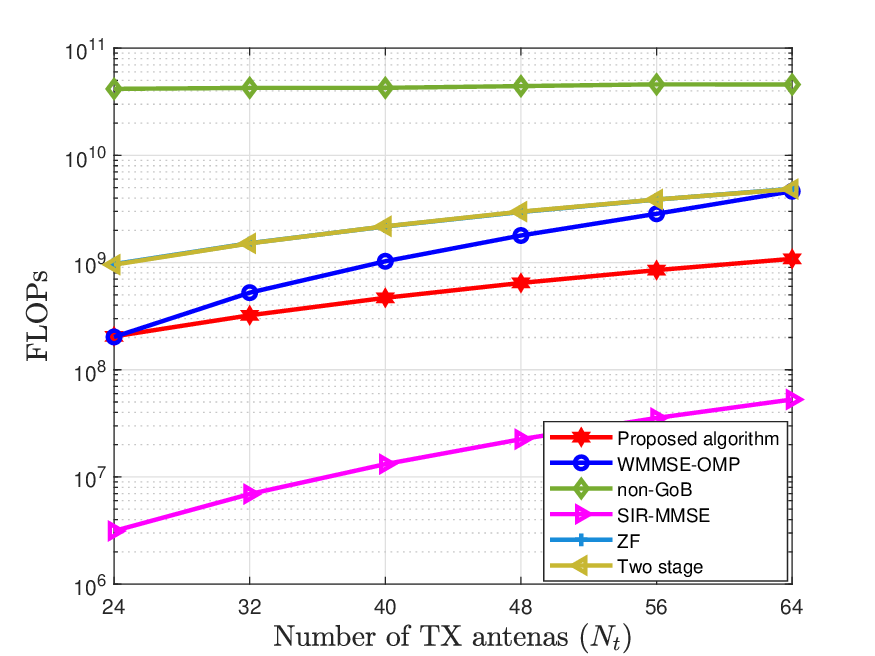}
    \caption{FLOPs vs. the number of transmit antennas at each O-RU when $P_{r} = 30$ dBm, $\Pi_{r,u} = 2$, $FL_{r} = 8$ Gbps, $R = 4$, $K = 4$, $M_{t} = M_{r} = 8$, and $N_{r} = 12$.}
    \label{Fig:FLOPs_vs_Nt2}
    % \vspace{-5mm}
\end{figure}

Finally, this subsection discusses the complexity of all the schemes in terms of both FLOPs and simulation runtime.
\cref{Fig:FLOPs_vs_Nt2} compares the computational complexity of the proposed algorithm with that of the benchmark schemes. The non-GoB scheme exhibits the highest complexity due to its reliance on CVX-based optimization of the digital precoder, whereas the SIR–MMSE scheme has the lowest complexity owing to its non-iterative digital precoder design. The proposed algorithm exhibits moderate complexity, surpassed only by the SIR–MMSE scheme. Given its previously demonstrated superior SE performance, the proposed algorithm offers the most favorable trade-off between performance and complexity among all considered schemes.

\begin{figure}
    \centering
    \includegraphics[width = \figW\textwidth]{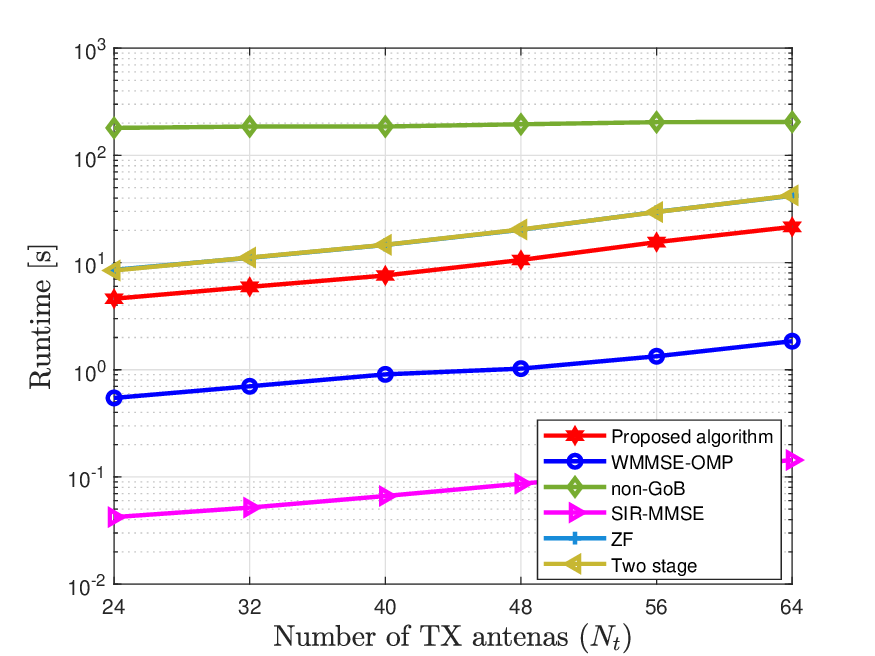}
    \caption{Runtime vs. the number of transmit antennas at each O-RU when $P_{r} = 30$ dBm, $\Pi_{r,u} = 2$, $FL_{r} = 8$ Gbps, $R = 4$, $K = 4$, $M_{t} = M_{r} = 8$, and $N_{r} = 12$.}
    \label{Fig:Runtime_vs_Nt}
    % \vspace{-5mm}
\end{figure}

The wall-clock runtime of all the schemes is presented in \cref{Fig:Runtime_vs_Nt}. These results are generated using MATLAB R2025a on Windows 11 Pro 25H2, running on a computer with 64 GB of RAM and an Intel(R) Core(TM) Ultra 7 265K (3.90 GHz) processor. It is noted that all the schemes, except non-GoB, show increasing runtime with the number of antennas. In contrast to the results in the previous figure, WMMSE-OMP achieves lower runtime than the proposed algorithm. 
Our investigation indicates that a substantial portion of the runtime of the proposed algorithm is attributable to the non-contiguous memory access and allocation overhead associated with its implementation, particularly within the MO subroutine used to design the analog precoders and combiners. In addition, the proposed algorithm is implemented in a modular manner, with its core functionalities distributed across multiple functions and files, which introduces additional function-call and memory-access overhead during execution. In comparison, WMMSE-OMP is implemented more compactly, and its analog precoder and combiner design primarily involves selecting vectors from a finite-size codebook. These implementation and algorithmic differences collectively contribute to the lower wall-clock runtime observed for WMMSE-OMP.

\section{Conclusion} \label{Sect:Concl}
This work investigated beamforming strategies for an O-RAN-based hybrid MIMO system employing fully connected hybrid architectures at both the BS and the users. In accordance with O-RAN specifications, the BS is disaggregated into multiple O-RUs, each equipped with a hybrid transceiver, a dedicated FHL to the O-DU, and an individual transmit power constraint. Beyond conventional constraints, such as unit-modulus requirements on analog beamformers and per-O-RU transmit power limits, the presence of FHLs introduces additional constraints that require the designed parameters to remain within the capacity limits of each O-RU’s FHL.

Under these constraints, analog and digital beamformers are jointly designed to maximize network sum SE while maintaining low computational complexity. Specifically, analog beamformers at both the O-RUs and users are optimized using gradient-based methods. While digital precoders are designed using the Lagrange multiplier method, the digital combiners are designed to ensure the received streams at the users are parallel to each other. Unlike most benchmark schemes that rely on EVDs or SVDs, the proposed approach avoids such operations without sacrificing performance, thereby reducing computational complexity.

Performance evaluation demonstrates that the proposed algorithm consistently achieves the highest SE among state-of-the-art benchmark schemes. Although its computational complexity falls behind one benchmark scheme, the proposed algorithm offers the best overall trade-off between performance and complexity, delivering high SE reliability and improved energy efficiency.

We note that validating the proposed algorithm in an actual O-RAN environment, and extending it to, for example, subarray-based HMSs and the max-min problem, constitutes an interesting future research direction.

\appendices
\section{Rationale for using (\ref{eq:FLr_scaling}) to comply with FHL constraint} \label{appendix:why_FLr_scaling}
    Suppose that the following system and network parameters are given: air-interface SE = 16 bps/Hz, $BW = 1$ Hz, $R_{code} = R_{use}$ = 1, ${\Pi_{r,u}} = 4$, $R=1$, $\setK{r} = 1$, $Q_{r} = O_{r} = 2$ bits, $t_{slot} = 1$ s, and $FL_{r} = 20$ bps. With these parameters, the O-DU can use any modulation scheme with 4 bits/symbol, e.g., 16-QAM, to represent each data symbol, while satisfying both the FHL capacity constraint and air-interface SE.
    Now, consider a tighter FHL constraint with $FL_{r} = 16$ bps. One possible approach is to reduce the modulation order to 8-QAM. However, reducing the modulation order alone may result in the O-RU maintaining a transmit power level designed to support the original air-interface conditions, despite the reduced amount of payload that can be delivered at that power level.
    
    Because the tighter $FL_{r}$ constraint necessitates a reduction in the modulation order and, consequently, the payload, the proposed scaling operation in (\ref{eq:FLr_scaling}) further reduces the O-RU transmit power so that the air-interface SE is reduced to the minimum level required (e.g., 12 bps/Hz) to accommodate the new reduced payload. This reduction in SE decreases the corresponding data volume transmitted over the FHL and enables the FHL capacity constraint to be satisfied while avoiding unnecessarily high transmit power. Consequently, the proposed operation can improve the system energy efficiency (EE) under tight FHL constraints.

\printbibliography

% end of the document
\end{document}